%% file: main.tex
\documentclass[journal=jpcbfk,manuscript=article]{achemso}

\usepackage[version=3]{mhchem} 
\usepackage{siunitx} 
\DeclareSIUnit\kcal{kcal}
\DeclareSIUnit\angstrom{\text {Å}}

\usepackage{xcolor}
\usepackage{subfig}

\newcommand{\gt}{>}
\newcommand{\lt}{<}

\author{Leo Christanell}
\affiliation[LMU]
{Department of Chemistry, Ludwig-Maximilians-Universität München, München, 81377, Germany}
\alsoaffiliation[CeNS]{Center for NanoScience (CeNS), Schellingstr. 4, München, 80799, Germany}

\author{Karl-Jakob König}
\affiliation[LMU]
{Department of Chemistry, Ludwig-Maximilians-Universität München, München, 81377, Germany}

\author{Julian Holzinger}
\affiliation[LMU]
{Department of Chemistry, Ludwig-Maximilians-Universität München, München, 81377, Germany}

\author{Anne K. Schütz}
\affiliation[LMU]
{Department of Chemistry, Ludwig-Maximilians-Universität München, München, 81377, Germany}

\author{Benjamin P. Fingerhut}
\email{benjamin.fingerhut@cup.lmu.de}
\affiliation[LMU]
{Department of Chemistry, Ludwig-Maximilians-Universität München, München, 81377, Germany}
\alsoaffiliation[CeNS]{Center for NanoScience (CeNS), Schellingstr. 4, München, 80799, Germany}

\title[DMP-ion interactions]
  {Molecular origin of $^{31}$P-NMR chemical shifts of phosphate groups with bivalent counter ions}
  
\abbreviations{IR,NMR,UV}

\begin{document}


%
%
%


\begin{abstract}

The electrostatic interactions of 
phosphate groups and counter ions critically affect the structure, function and reactivity of DNA or RNA.
We present a joint experimental-theoretical investigation of dimethyl phosphate 
(\ce{DMP^-}) in aqueous solution, an established model system of the sugar-phosphate backbone.
Utilizing $^{31}$P-NMR spectroscopy as probe of phosphate-ion association, 
variations of \ce{Mg^{2+}} and \ce{Ca^{2+}} content exhibit a systematic shielding 
of  the $^{31}$P chemical shift ($\delta_{iso}$($^{31}$P)) with moderate temperature dependence.
Enhanced sampling molecular dynamics (MD)
and ab initio (GIAO-DF-LMP2) level of theory are used to reveal the microscopic mechanism.
Simulations 
 are performed for a configurational ensemble of \ce{DMP^-}-ion geometries and their first solvation shells,  demonstrating (i) the spatial convergence of changes of the nuclear shielding constant $\sigma_{iso}$($^{31}$P),
 (ii) the intramolecular geometric origin of short-timescale $\sigma_{iso}$($^{31}$P) 
fluctuations  and (iii) an average 
shift of $\sigma_{iso}$($^{31}$P)
of about 3-5 ppm upon contact ion pair formation with  Mg$^{2+}$ or Ca$^{2+}$ ions. 
A quantitative analysis of $\delta_{iso}$($^{31}$P)
for varying ion 
content and temperature  
allows us to extract the temperature-dependent fraction of the contact ion pair species, indicating that solvent separated or free ion pairs are the energetically preferred species.
The results 
impose boundary conditions for improvements of phosphate 
ion force fields and establish the interactions underlying the changes of $\delta_{iso}$($^{31}$P).


 \end{abstract}

\section{Introduction}




The polyanionic structures of DNA and RNA are stabilized via electrostatic interactions with counter ions.\cite{Wong:AnRevPhysChem:2010,Lipfert:AnnRevBiochem:2014}
In this context, bivalent counter ions like Mg$^{2+}$
are particularly important for neutralizing the negative charge density of the phosphate (PO$_2$) groups, thereby facilitating the folding of complex RNA structures.\cite{Draper:AnnRevBiophys:2005,Denesyuk:NatChem:2015} 
Moreover, metal ions are directly involved in the catalytic function of RNA.\cite{Cate:NatStructBiol:1997,Inoue:NucAcidRes:2004,Schnabl:CurOpChemBiol:2010}
Charge neutralization of DNA or RNA occurs on a length scale of about 2 nm  from surface.\cite{Laage:ChemRev:2017,Fingerhut:ChemCommun:2021} It gives rise to an ion atmosphere around the polyanion that is made up from 
the distinct species of contact ion pairs (CIP) and solvent shared ion pairs (SSIP). Both species form specific hydrogen bond interactions, 
whereas weakly bound ions interact via
\emph{unspecific} (Coulombic) interactions. 
In CIPs the cation is partially dehydrated and interacts  directly with one of the non-bridging oxygens of the PO$_2$ group (inner-sphere coordination), while in SSIPs the cation retains its full solvation shell (outer-sphere coordination).
In RNA about 10\% of bivalent ions are estimated to be site specifically bound.\cite{Freisinger:CoordChemRev:2007} 

Ion counting experiments \cite{Bai:JACS:2007,Trachman:NucAcidRes:2017,Jacobson:NucAcidRes:2016,Pabit:JACS:2010} 
provide quantitative access to the total number of ions interacting with the polymer, demonstrating that \ce{Mg^{2+}} binds stronger to RNA than DNA,\cite{Gebala:BiophysJ:2019,Pollack:AnnRevBiophys:2011} without the ability to distinguish specific interacting species, i.e., CIP and SSIP on the molecular level.
Recently, infrared (IR) and non-linear two-dimensional IR (2D-IR) spectroscopy together with in-depth theoretical analysis suggested the presence of CIP species for double-stranded RNA~\cite{Fingerhut:BiophysJ:2021} and transfer RNA (tRNA)~\cite{Schauss:JPCB:2021},
and a molecular electrostatic scenario at the surface of the biomolecules not reproduced by Poisson-Boltzmann theory.
CIP of \ce{Mg^{2+}} with PO$_2$ groups have the ability to strongly modulate the local electrostatic potential, thereby facilitating the close approach of PO$_2$ groups in kink and bulge regions of folded RNA.
 The specific detection of CIP in the biomolecules relied on a characteristic blue-shift of the asymmetric  
PO$_2$ stretching vibration, that showed a number of CIP consistent with fluorescence titration experiments. 
Analogous observations of a blue-shifted asymmetric  PO$_2$ stretching vibration for
ionic (\ce{Na^{+}}, \ce{Mg^{2+}}, \ce{Ca^{2+}}) solutions of \ce{DMP-},~\cite{Schauss:JPCL:2019,Schauss:JPCL:2019_2,Fingerhut:ZPhysChem:2020}
 a well established model system of the sugar-phosphate backbone of DNA and RNA, 
\cite{Florian:JACS:98,Fingerhut:JCP:2015,Puyo-Fourtine:JPCB:2022,Delgado:JCIM:24}
allowed for rigorous simulation access to the origin of the blue shift of the vibration. 
Microscopically, the close proximity of the Mg$^{2+}$ ion and the non-bridging oxygen atom of the PO$_2$ group (P-O$^-$ - \ce{Mg^{2+}}  distance $\approx$ 2.1 \AA) leads to an asymmetric distortion of the PO$_2$ bonding potential due to exchange repulsion interactions at short inter-molecular distances.
Estimates utilizing the amplitudes of  IR and 2D-IR spectroscopy  suggested the formation of CIP species in a 10-30~\% range for 10 equivalents (eq.) excess ion concentration relative to the 0.2 M concentration of \ce{DMP-}.
%
%
%
%
%
Kutus and coworkers recently derived ion pairing equilibria of \ce{Na+}, \ce{Mg^2+} and \ce{Ca^2+} with \ce{DMP-} from dielectric relaxation spectroscopy.\cite{Kutus:JMolLiqu:22}
Respective association constants relied on estimates of dipole moments of dipole-bound complexes from density functional theory and allowed to derive ion pair lifetimes of 330 ps for \ce{Na+}\ce{DMP-} and 700 ps for \ce{Ca^2+}\ce{DMP-}, 
 while \ce{Mg^2+} \ce{DMP-} lifetime estimates were consistent with $>$ 1 $\mu$s estimates.

Utilizing the spin 1/2 $^{31}$P nucleus of the phosphate group, NMR spectroscopy has long been used for detecting metal ions, in particular \ce{Mg^2+} interactions with DNA or RNA \cite{Gueron:PNAS:75,Dinh:NAR:77,Son:EurJBiochem:1977,Rose:PNAS:80}
and to rank \ce{Mg^2+} binding sites in the polyphosphate chain.\cite{Son:EurJBiochem:1977}
Haake and coworkers \cite{Haake:InorChem:84} suggested a two-component model to extract association constants, by utilizing the $^{31}$P 
chemical shift with increasing ion concentration, the latter being tentatively assigned to ion pairing.
A smilar model yielded association constants of \ce{DMP^-} and ions that are largely consistent between $^{31}$P-NMR and dielectric relaxation spectroscopy.\cite{Kutus:JMolLiqu:22}
%
Besides these quantitative approaches to ion pairing, NMR experiments characterized 
numerous ion binding sites.\cite{Donghi:book:2012,Marino:JACS:94,Fuertig:ChemBioChem:2003}
Nevertheless, the quantitative interpretation of observed  $^{31}$P chemical shifts 
is still complicated by a limited microscopic understanding of their origin,
in particular the dependence of chemical shifts on sugar-phosphate conformation and sensitivity to ionic interactions.
Moving beyond structure to chemical reactivity, in particular phosphoryl cleavage and transfer reactions, a deeper understanding of the origin of $^{31}$P  chemical shifts could elucidate biocatalytic mechanisms.\cite{Szanto:JCTC:2024,Shein:NatChem:2024}

Concerning the association of PO$_2$ groups with bivalent cations \ce{Mg^2+} and \ce{Ca^2+}, substantial uncertainty remains regarding their tendency to form CIP structures.
Ion association constants from dielectic relaxation spectroscopy suggest largely similar properties for \ce{Mg^2+} and \ce{Ca^2+} \cite{Kutus:JMolLiqu:22}.
MD simulations with a fixed-charge force field, utilizing ion parameters by Li \emph{et al.}~\cite{Li:JCTC:2013},
suggest a stronger tendency for \ce{Mg^2+} than \ce{Ca^2+} to compress and assemble DNA strands into compact structures.\cite{Xu:JMolLiqu:21}
A recent re-parametrization of the 12-6-4 non-bonding model by the Merz group\cite{Findik:JCTC:24}, based on ion binding free energy energies for dihydogen-phosphate (\ce{H_2PO4^-}), suggests a  
stronger binding of \ce{Mg^2+} to phosphates, compared to \ce{Ca^2+}.
For both bivalent ions, CIP are substantially more stable than SSIP.
Compared to \ce{DMP^-}, the asymmetric  
PO$_2$ stretching vibration of \ce{H_2PO4^-} has a substantially more complex vibrational structure,\cite{Costard:JCP:2015,Kundu:JCP:2024} where water molecules in the first solvation shell 
donate hydrogen bonds to the PO$_2$ moiety and accept hydrogen  bonds from the O-H groups of \ce{H_2PO4^-}, a scenario not realized in \ce{DMP^-} amd to which extent the hydrogen bond donor functionality\cite{Findik:JCTC:24} of \ce{H_2PO4^-} affects the relative stability of CIP and SSIP is an open question.


A recent re-parametrization 
of \ce{DMP^-} with the polarizable amoeba model\cite{Puyo-Fourtine:JPCB:2022}
provided reasonable values of ion binding free energy energies  of \ce{DMP-} with \ce{Mg^2+} and \ce{Ca^2+}, respectively.
Nevertheless, the refined force field tends to underestimate the population of CIP,  compared to estimates derived from 2D-IR experiments, indicating an over-stabilization of the SSIP species. 
To the contrary, theoretical work utilizing the  reference interaction site model suggests that \ce{Ca^2+} ions have a preference for CIP formation over  \ce{Mg^2+} ions.\cite{Nguyen:JPCB:2020}
%
%
 Force fields used in MD simulations of ion association around DNA and RNA thus still have limited predictive accuracy in describing the interactions of phosphate 
groups with doubly charged ions.
Further experimental benchmark data on the relative stability of CIP and SSIP species and the behaviour concerning the differences of bivalent ions are highly desirable.

In this work 
we present the temperature dependence of $^{31}$P chemical shifts in \ce{DMP^-} for a wide range of  \ce{Mg^2+}  concentrations and compare to measurements with other ions (\ce{Ca^2+}, \ce{Zn^2+}, \ce{Cd^2+}, \ce{Na^+} and \ce{[Co(NH_3)_6]^{3+}}).
The observations are analyzed in ab initio simulations investigating the convergence of the quantum mechanical interaction region and basis set dependence.   
 The origin of the  observed chemical shift is microscopically assigned to the formation of the CIP species, which is distinct from the effect of geometric distortions of \ce{DMP^-}. 
 Utilizing a two-component model, we further establish the relative population $X[CIP]$ of CIP with \ce{Mg^2+} and \ce{Ca^2+} ions, which in turn imposes boundary conditions on the potential of mean force of ion binding.
\section{Materials and Methods}

\subsection{NMR spectroscopy}

\label{sec:NMRexp}
\input{nmr_methods/nmr_methods_2}

\begin{table}
  \caption{Summary of the metal ion stock solutions used in the NMR experiments.}
  \label{tbl:sample_prepartation}
  \begin{tabular}{lll}
    \hline
    Metal ion & Salt & Stock concentration / M \\
    \hline
    \ce{Mg^2+} & \ce{MgCl2} & 4.07 \\
    \ce{Ca^2+} & \ce{CaCl2} & 6.00 \\
    \ce{Cd^2+} & \ce{CdI2} & 0.50 \\
    \ce{[Co(NH3)6]^3+} & \ce{[Co(NH3)6]Cl3} & 0.25 \\
    \ce{Zn^2+} & \ce{ZnCl2} & 3.57 \\
    \ce{Na^+} & \ce{NaCl} & 4.94 \\
    \hline
  \end{tabular}
\end{table}

\subsection{Simulations}

\subsubsection{Molecular Dynamics Simulations}
\input{tc_methods/md_methods}

\subsubsection{Ab Initio Simulations}
\input{tc_methods/qm_methods}

\section{Results}

\subsection{Concentration and temperature dependence of $^{31}$P chemical shifts}

\begin{figure}
  \includegraphics[width=\textwidth]{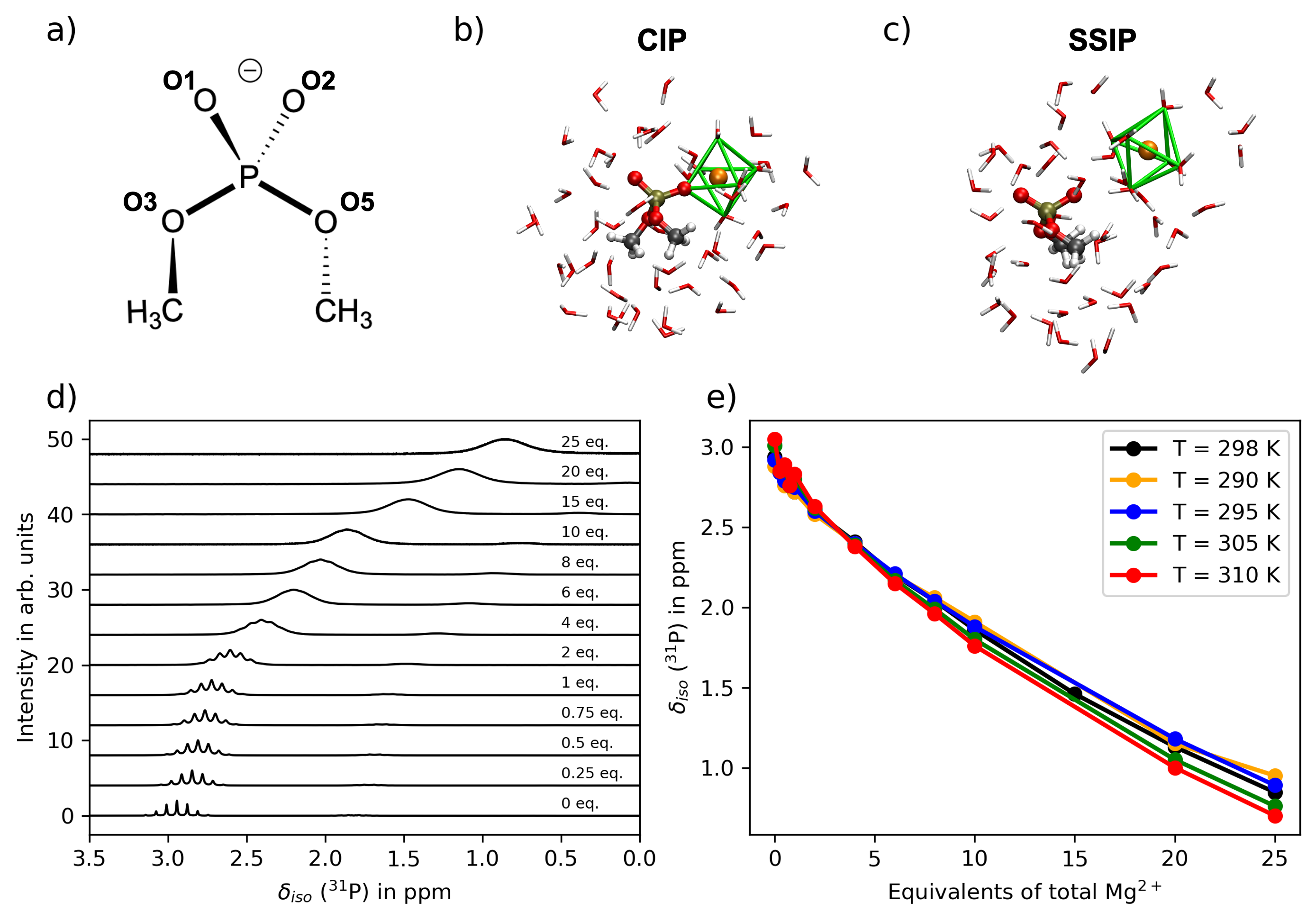}
  \caption{Experimental $\delta_{iso}$($^{31}$P) of \ce{DMP-} upon titration with \ce{Mg^2+}. (a) Atomic labeling of \ce{DMP-}. 
(b,c) Snapshots of \ce{DMP-} - \ce{Mg^{2+}} CIP and SSIP from MD simulations. The octahedral solvation shell of \ce{Mg^2+} is highlighted in green, i.e., \ce{Mg}O1(\ce{H2O})$_5$ and \ce{Mg}(\ce{H2O})$_6$.
  (d) Experimental \ce{^31P}-NMR spectra measured in the range \SIrange{0}{25}{} eq. of \ce{Mg^{2+}} at \SI{298}{\kelvin}. (e) Peak positions of $\delta_{iso}$($^{31}$P)
  in a temperature range from \SIrange{290}{310} {\kelvin} and \SIrange{0}{25}{} eq. of \ce{Mg^2+}.}
  \label{fgr:figure_1}
\end{figure}

A central question concerns the sensitivity of $^{31}$P NMR observable to detect in a species selective manner CIP or SSIP of the PO$_2$ group with bivalent ions. 
Observed changes of $^{31}$P chemical shifts $\delta$($^{31}$P)
arise due to the high sensitivity of the 1/2 spin nucleus to the local molecular environment that is affected by hydration species, the \ce{DMP-} conformation and electronic density changes from specific interactions of ions with the PO$_2$ group
(exemplary structures of \ce{DMP-} - \ce{Mg^{2+}} CIP and SSIP are depicted in Figure~\ref{fgr:figure_1},b-c). We measured the isotropic chemical shift of the  $^{31}$P nucleus ($\delta_{iso}$($^{31}$P)) for 0.2 M NaDMP dissolved in water 
as a function of counter ion concentration and temperature.
Under such low concentration condition of \ce{Na+} \ce{DMP-} association is minor\cite{Kutus:JMolLiqu:22} and the effect on  $\delta_{iso}$($^{31}$P) is negligible.
%
%

Figure~\ref{fgr:figure_1},e presents  the $\delta_{iso}$($^{31}$P) 
for varying \ce{Mg^2+} concentration  
in the range \SIrange{0}{25}{}~eq. at \SI{298}{\kelvin} without $^1$H decoupling. 
%
At low \ce{Mg^2+} content ($\leq$ 2 eq.), the septet is caused by the $^3$J$_{PH}$ coupling of the  $^{31}$P spin to the six magnetic equivalent $^1$H. $\delta_{iso}$($^{31}$P) peak positions derived from  $^1$H-decoupled and $^1$H non-decoupled NMR spectra (see Materials and Methods Section) essentially coincide.
For \ce{Mg^2+} content $>$ 4 eq., a broadened unstructured  \ce{^31P} resonance is observed, which is increasingly shielded with a further increase of \ce{Mg^2+}.
Changes in the linewidth of the \ce{^31P}-NMR spectra are associated either with 
changes of the transverse relaxation rate $R_2$ or with chemical exchange (e.g. \ce{Mg^{2+}}-bound and free \ce{DMP-}) or a combination of both (cf. Table~\ref{tbl:exp}).


%
%

Figure~\ref{fgr:figure_1},d depicts the peak position of $\delta_{iso}$(\ce{^31P}) for varying \ce{Mg^2+} concentration and temperature in a range  \SIrange[range-phrase=\,--\,, range-units=single]{290}{310}{\kelvin}.
The overall range of $\delta_{iso}$(\ce{^31P})  covers a
 shift of about \SI{2.5}{ppm} that is moderately increased ($\approx$ \SI{0.4}{ppm}) with increasing temperature and 
$\delta_{iso}$(\ce{^31P})  does not approach saturation, even for 25~eq. of \ce{Mg^2+}. 


Table~\ref{tbl:exp} summarizes the experimental results for ions \ce{Mg^{2+}}, \ce{Ca^{2+}}, \ce{Na^{+}}, \ce{Zn^{2+}}, \ce{Cd^{2+}}, \ce{[Co(NH_3)_6]^{3+}}.
Major changes of $\delta_{iso}$(\ce{^31P}) are observed for \ce{Mg^{2+}}, \ce{Ca^{2+}} and 
$^[$\footnote{Note that difficulties from strong linebroadening for \ce{Zn^2+} limited the analysis to 
 $<$\SI{0.1}{eq}}$^]$\ce{Zn^{2+}}
 while for  \ce{Cd^{2+}}, \ce{[Co(NH_3)_6]^{3+}} and \ce{Na^{+}} changes of $\delta_{iso}$(\ce{^31P}) are smaller by about an order of magnitude.
Similar trends are observed for the longitudinal relaxation rate $R_1$ and transverse relaxation rate $R_2$ where \ce{Mg^{2+}}
and \ce{Zn^{2+}} show changes that are substantially larger than that for 
\ce{Cd^{2+}}, \ce{[Co(NH_3)_6]^{3+}} and \ce{Na^{+}}, 
correspondingly reflected in the changes of the linewidth.

\begin{table}
  \caption{$^{31}$P-NMR observables for 0 eq. (reference) and 4 eq. of the respective ion at \SI{298}{\kelvin}. $R_1$ denotes the longitudinal relaxation rate and $R_2$ the transverse relaxation rate, $\Delta\delta_{\text{iso}}$ is the change in chemical shift with respect to 0 eq. of the ions.
\ce{Zn^2+} is indicated for 0.1 eq. as data analysis at higher concentrations was prevented by strong line broadening. The observed $R_2$ relaxation rate is the sum of intrinsic relaxation rate plus a possible exchange contribution.}
  \label{tbl:exp}
  \begin{tabular}{llllll}
    \hline
    Metal ion & $\delta_{\text{iso}}$ / ppm & $R_1$ /  \si{\hertz} &  $R_2$ /  \si{\hertz}  & linewidth / \si{\hertz} &$\Delta\delta_{\text{iso}}$ / ppm\\
    \hline
    0 eq. (ref.)                &  2.94  &  0.09   & 0.28   & 0.40 & -- \\
    \ce{Mg^2+}        &  2.41  &  0.19    & 27.78    & 14.85     & -0.53      \\
    \ce{Ca^2+}        &  2.57  &  0.11    & 0.71   & 1.55 & -0.37      \\
    \ce{Cd^2+}        &  2.98  &  0.09    & 0.57   & 2.64 & 0.04       \\
    \ce{[Co(NH3)6]^3+}&  2.90  &  0.09    & 0.96   & 1.98 & -0.04      \\
    \ce{Zn^2+}        & 2.92   &  0.33    & 69.93    & 25.54 & -0.02       \\
    \ce{Na^+}         & 3.01   &  0.09    & 0.33   & 2.98 & 0.07     \\
    \hline
  \end{tabular}
\end{table}

\subsection{Solvent shell and geometric dependence of the \ce{^31P} chemical shift}

\subsubsection{Spatial interaction range}   
To obtain a microscopic understanding of the observed 
shift of $\delta_{iso}$(\ce{^31P}) 
with increasing \ce{Mg^2+} content, we investigated in ab initio simulations the convergence of the isotropic \ce{^31P} shielding constant ($\sigma_{iso}$(\ce{^31P})) for a quantum mechanical interaction range of increasing size.
Experimentally measured $\delta_{iso}$(\ce{^31P}) are related to the microscopic quantity $\sigma_{iso}$(\ce{^31P}) as\cite{Rusakova:IJMS:2026} 
\begin{equation}
\delta_{iso}(^{31}P) \approx  \sigma_{iso}^{Ref}(^{31}P) - \sigma_{iso}(^{31}P) 
\end{equation}
where in experiments phosphoric acid is used as reference.
The hydration environment of a \ce{DMP-} - \ce{Mg^{2+}} CIP was sampled from a classical molecular dynamics trajectory, taking into account \ce{DMP-} - \ce{Mg^2+} $\times$ N(\ce{H2O}) clusters of increasing size, with the largest clusters containing 60 water molecules (\SI{5.5}{\angstrom} water shell, see Material and Methods Section), required to form about three hydration shells around the CIP.
The results of the \ce{DMP-} - \ce{Mg^{2+}} CIP are compared to \ce{DMP-}$\times$ N(\ce{H2O}) clusters, describing the hydration environment of free \ce{DMP-} in aqueous solution.
%

\begin{figure}
 \includegraphics[width=0.8\textwidth]{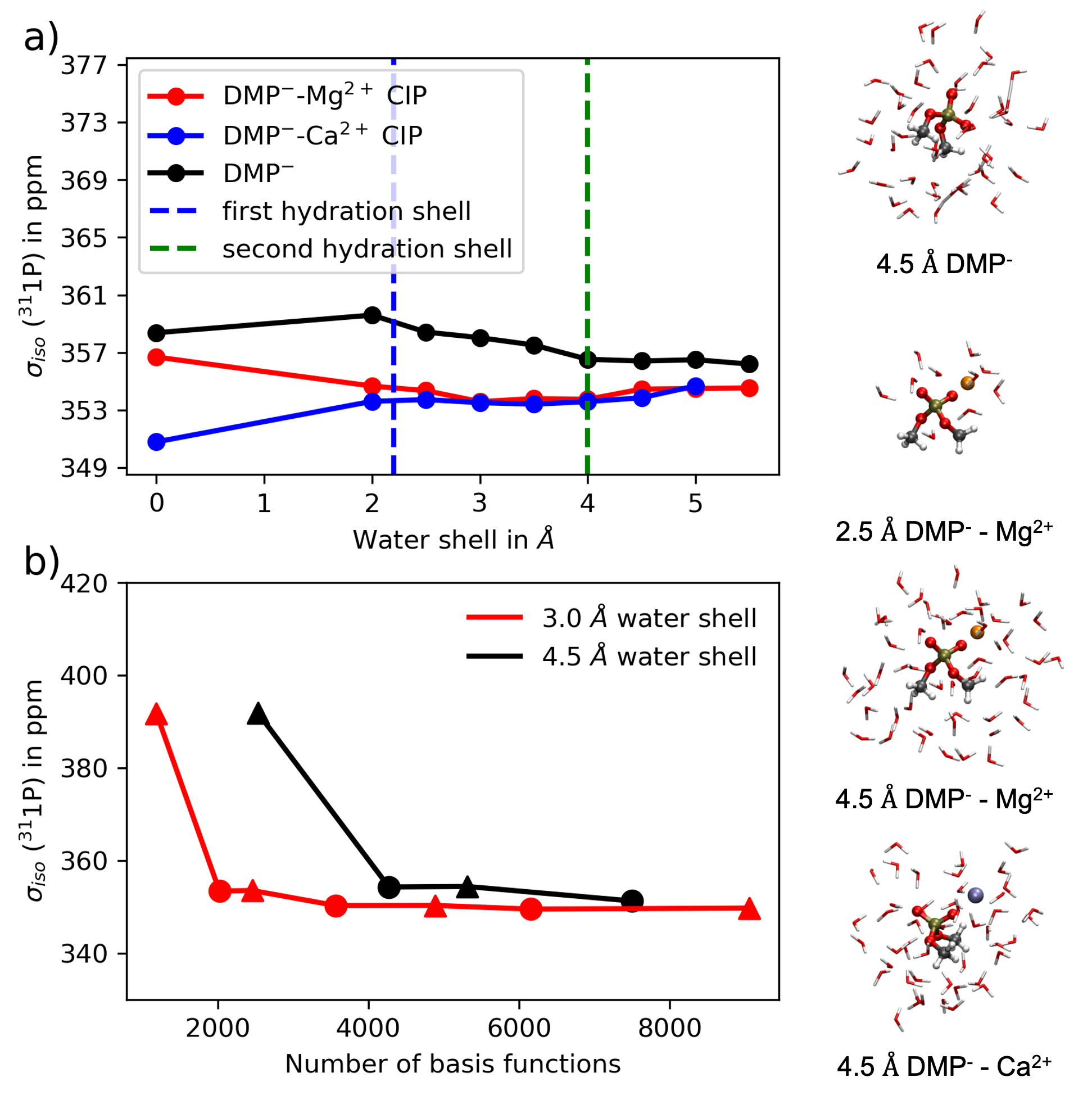}
  \caption{ 
  Benchmark of hydration shell and basis set. (a) Variation of $\sigma_{iso}$($^{31}$P) in a randomly chosen \ce{DMP-} - \ce{Mg^2+} CIP configuration (red), \ce{DMP-} - \ce{Ca^2+} CIP configuration (blue) and free \ce{DMP-} (black) with increasing size of the hydration shell (basis: pCVTZ). Dashed vertical lines indicate the first and second hydration shell, consisting of 6 and 36 \ce{H_2O} (\ce{DMP-} - \ce{Mg^2+} CIP); the largest cluster with \SI{5}{\angstrom}~water shell  considers  60 \ce{H_2O}.
(b) Benchmark of $\sigma_{iso}$($^{31}$P) in the \ce{DMP-} - \ce{Mg^2+} CIP  for varying basis size pCVXZ (X=D, T, Q, 5) basis set with a \SI{3.0}{\angstrom} (red) and \SI{4.5}{\angstrom} water shell (black). Triangles indicate a uniform pCVXZ basis set for all atom types and circles indicate a mixed basis set with pVDZ employed for hydrogen atoms. 
The right panel shows \ce{DMP^-} $\times$ N \ce{H_2O} and \ce{DMP-} - \ce{I^2+} CIP $\times$  N \ce{H_2O}  (I = Mg, Ca) cluster snapshots for hydration shells of varying size. 
  }
  \label{fgr:figure_2}
\end{figure}

Figure~\ref{fgr:figure_2},a depicts $\sigma_{iso}$(\ce{^31P}) for an increasing solvation shell around the \ce{DMP-} - \ce{Mg^{2+}} CIP. Initially, the addition of a few discrete \ce{H2O} molecules decreases $\sigma_{iso}$(\ce{^31P}) by about \SIrange[range-phrase=\,--\,, range-units=single]{2}{3}{ppm} (\SIrange[range-phrase=\,--\,, range-units=single]{2}{3}{\angstrom} solvation  shell). Due to \ce{H2O}  belonging to the second and third solvation shell, $\sigma_{iso}$ is moderately increased by about \SI{1}{ppm}, leading to convergence of $\sigma_{iso}$($^{31}$P) on a spatial \SIrange[range-phrase=\,--\,, range-units=single]{4.5}{5}{\angstrom} range, where a complete second and third solvation shell, containing about 50-60 \ce{H2O} molecules, are formed around the CIP.
In comparison, the hydration environment around free \ce{DMP-} (black dots in Fig.~\ref{fgr:figure_2},a) first induces a shielding by about \SI{2}{ppm} from the addition of discrete water molecules that converges on a very similar spatial range of \SIrange[range-phrase=\,--\,, range-units=single]{4.5}{5}{\angstrom}, i.e., for a hydration environment consisting of three solvation shells.
The results demonstrate, that $\sigma_{iso}$($^{31}$P) is particularly sensitive to details of the local environment: Changes of the charge density of the PO$_2$ group are induced by the first, second and third solvation shell, successively defining its polarization, 
which is reflected in \SIrange[range-phrase=\,--\,, range-units=single]{2}{3}{ppm} changes of $\sigma_{iso}$($^{31}$P).
The $^{31}$P nucleus thus acts as local probe for environmental changes imprinted on the charge density of the  PO$_2$ group, limited to a $<$ \SI{5}{\angstrom} range.
%

Additionally, we explored the effect of size of the basis set on simulated $\sigma_{iso}$($^{31}$P) in a \ce{DMP-} -  \ce{Mg^2+} CIP with a 3.0 and \SI{4.5}{\angstrom} water shell, respectively (Fig.~\ref{fgr:figure_2},b).
While there is a substantial influence of basis set size (up to \SI{40}{ppm}), when going from double zeta 
to triple zeta 
quality, changes are more moderate when comparing the results of triple and quadruple zeta (triangles in Fig.~\ref{fgr:figure_2},a).  Comparison of quintuple and quadruple basis set for the \ce{DMP-} - \ce{Mg^2+} CIPs with  a \SI{3.0}{\angstrom} hydration shell demonstrates that the quadruple zeta basis set yields essentially converged results.
%

A mixed-valence basis set has the potential to reduce the numerical effort, especially for \ce{DMP-} - \ce{Mg^2+} CIPs with large (\SI{4.5} {\angstrom}) hydration shell, where triple and quadruple zeta basis sets correspond to $\approx$ 4300 and 7500 basis functions, respectively. 
We find that reducing the basis to pVDZ quality for hydrogen atoms 
has a minor effect on $\sigma_{iso}$($^{31}$P)(compare triangles and dots in Fig.~\ref{fgr:figure_2},b). 
As the total basis set size can be substantially reduced (e.g. 4900 vs. 3600 basis functions for pCVQZ and pCVQZ/pVDZ , respectively), 
$^{31}$P  chemical shifts can be simulated rather efficiently with a mixed basis. Absolute errors by a mixed pCVQZ/pVDZ basis treatment are on the order of 0.5~ppm, 
comparable to the error or the GIAO-DF-LMP2 ab initio method.\cite{Loibl:JCP:2012} 

\subsubsection{Dependence of $\sigma_{iso}$($^{31}$P)  on intra-molecular degrees of freedom and \ce{DMP-} - \ce{Mg^2+} separation}

The above results demonstrate that reasonably efficient but accurate simulations of $\sigma_{iso}$($^{31}$P) can be achieved with a \SI{4.5}{\angstrom} solvation shell around CIP and using a mixed pCVTZ/pCVDZ basis for heavy elements carbon and phosphorous, and  hydrogen atoms, respectively. To investigate differences of $\sigma_{iso}$($^{31}$P) for  CIP, SSIP and fully solvated \ce{DMP^-} hydration geometries,
 we performed large scale sampling of hydration geometries along a reaction coordinate obtained by umbrella sampling MD simulations (see Simulation Details).
For each  \ce{DMP-} - \ce{Mg^2+} separation in the range [2-6]~\AA,~we sampled 100 independent solvent configurations (0.1~ns separation, 4.5~\AA~solvation shell, up to 60 \ce{H2O} molecules) 
and evaluated $\sigma_{iso}$($^{31}$P) along the reaction coordinate (2100 simulations total). 

\begin{figure}
  \includegraphics[width=\textwidth]{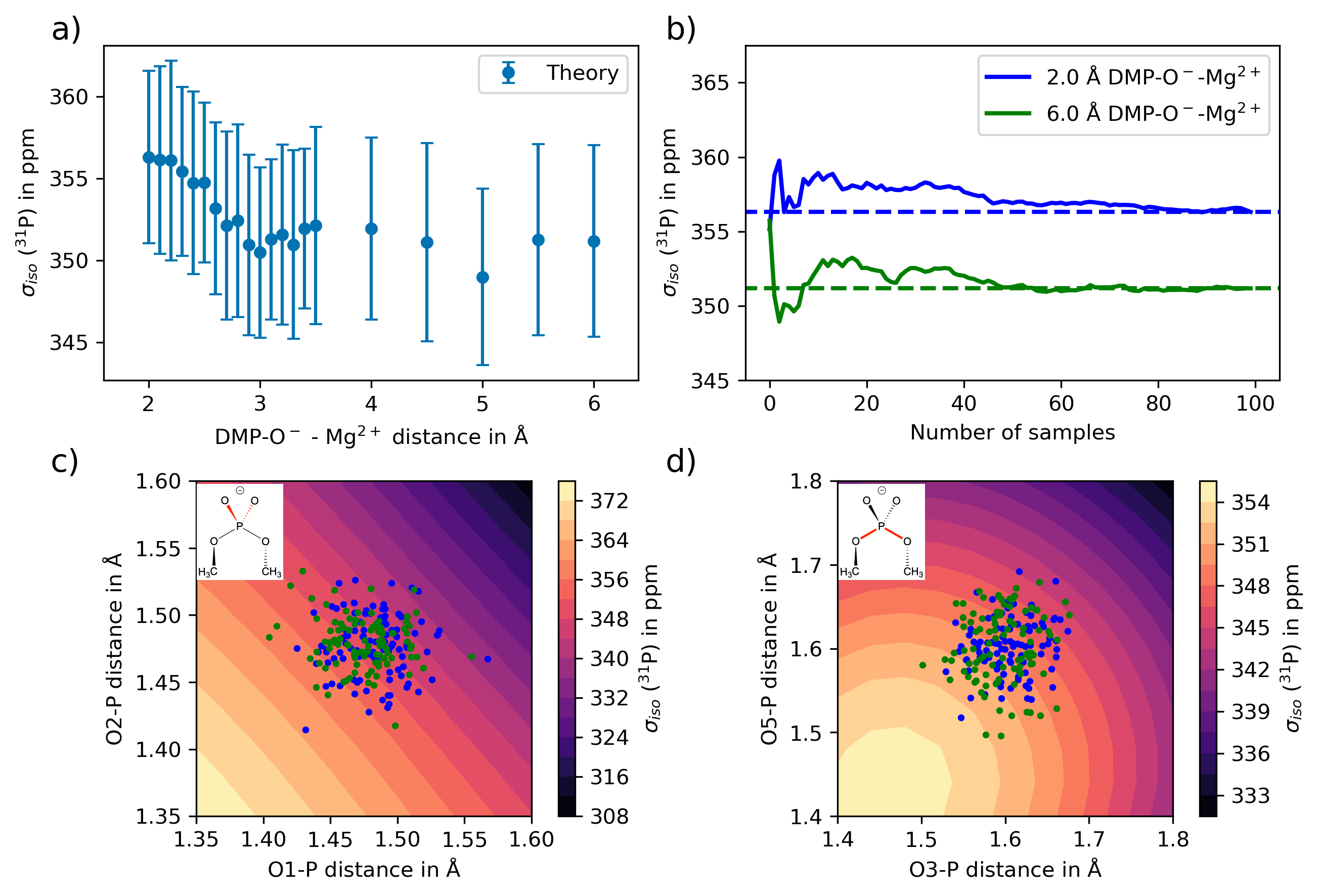}
  \caption{(a) Simulated isotropic \ce{^31P} shielding constants $\sigma_{iso}$($^{31}$P) as a function of \ce{DMP-} - \ce{Mg^2+} separation. 
  Averages (dots) over sampled  100 configurations are shown together with the standard deviation (bars). Configurations taken from umbrella sampling in the range \SIrange{2.00}{6.00}{\angstrom} (2100 simulations total) consider a \SI{4.5}{\angstrom} hydration shell (basis: pCVTZ/pVDZ).
  (b) Cumulative average of $\sigma_{iso}$($^{31}$P) over 100 configurations for two \ce{DMP-} - \ce{Mg^2+} separations (\SI{2.}{\angstrom} and  \SI{6.00}{\angstrom}). 
  (c) 2D surface of $\sigma_{iso}$($^{31}$P) for the variation of O1/O2-P bond lengths
  (constant O3/O5-P distance: \SI{1.60}{\angstrom}).
    Sampled O1/O2-P bond lengths in umbrella sampling MD simulations are indicated with blue (\ce{DMP-} - \ce{Mg^2+} separation: \SI{2.00}{\angstrom}) and green dots (\ce{DMP-} - \ce{Mg^2+} separation: \SI{6.00}{\angstrom}), respectively. 
 %
  (d) 2D surface of $\sigma_{iso}$($^{31}$P) for the variation of O3/O5-P bond lengths
  (constant O1/O2-P distance: \SI{1.48}{\angstrom}).
    Sampled O3/O5-P bond lengths in umbrella sampling MD simulations are indicated with blue (\ce{DMP-} - \ce{Mg^2+} separation: \SI{2.00}{\angstrom}) and green dots (\ce{DMP-} - \ce{Mg^2+} separation: \SI{6.00}{\angstrom}), respectively. 
}
  \label{fgr:figure_3}
\end{figure}

Figure~\ref{fgr:figure_3},a shows the mean (dots) and width (bars, standard deviation) of the $\sigma_{iso}$($^{31}$P) distributions as a function of \ce{DMP-} - \ce{Mg^2+} separation.
For all separations, the width of distributions covers about 10-12 ppm, much broader than the $\approx$ 0.2 ppm width observed in the NMR-experiments for \ce{Mg^2+} contents $\lt$ 2 eq (cf. Fig.~\ref{fgr:figure_1}).
The mean of the distributions shows a systematic shift from 356.3 ppm 
(\ce{DMP-} - \ce{Mg^2+} separation: 2~\AA),~ 
to 351.2 ppm 
(\ce{DMP-} - \ce{Mg^2+} separation: 6~\AA).
We have verified via the moving mean, that the center of respective $\sigma_{iso}$($^{31}$P) distributions is converged within \SI{0.2}{ppm} (cf. Fig.~\ref{fgr:figure_3},b).
%

Configurational sampling of the 
reaction coordinate of \ce{DMP-} - \ce{Mg^2+} separation employs an ensemble of instantaneous geometries with varying \ce{DMP-} bond lengths. 
 Microscopic understanding of the distributional widths was then obtained from the dependence of $\sigma_{iso}$($^{31}$P) on the intra-molecular degrees of freedom, the length of non-bridging  O1/O2-P bonds and bridging O3/O5-P bonds, as well as O1-P-O2 and O3-P-O5 bond angles and C-O3-O5-C dihedral angles (cf. inlays in Fig.~\ref{fgr:figure_3},b-d).
%
%
We find that $\sigma_{iso}$($^{31}$P) shows particular sensitivity to the O1/O2-P bond lengths and to a lower extend to the  O3/O5-P bond lengths ($\gt$ tens of ppm, see below), while the sensitivity on bond angles and the C-O3-O5-C dihedral angle is less pronounced ($\approx$\SI{4}{ppm} in the relevant angle distribution). 
%
Figure~\ref{fgr:figure_3},c-d present the dependence of $\sigma_{iso}$($^{31}$P) on respective bond lengths O1/O2-P and O3/O5-P.
The 2D-surfaces show a strong dependence on the particular \ce{DMP-} geometry with a monotonous decrease of $\sigma_{iso}$($^{31}$P) upon elongation of the bond lengths, spanning a $\gt$ \SI{50}{ppm} range for O1/O2-P distances and $\gt$ \SI{30}{ppm} range for O3/O5-P distances
(\ce{DMP-} equilibrium geometry: O1/O2-P bond length = 1.48~\AA, O3/O5-P bond length = 1.60~\AA). 
Comparison with the distribution of sampled bond lengths (blue and green points in Fig.~\ref{fgr:figure_3},c-d) shows that substantial width of the 2D-surfaces is explored due to the flexibility of \ce{DMP-} in MD simulations. 
As P=O and P-O bonds show oscillatory dynamics with characteristic frequencies in the \SIrange[range-phrase=\,--\,, range-units=single]{800}{1250}{\per\centi\meter} range\cite{Costard:JCP:2015}, corresponding to a timescale of a few tens of femtoseconds, the configurational sampling of instantaneous \ce{DMP-} geometries
induces a distributional width of $\sigma_{iso}$($^{31}$P) on the order of \SIrange[range-phrase=\,--\,, range-units=single]{15}{20}{ppm}.

The pronounced sensitivity of $\sigma_{iso}$($^{31}$P) on the \ce{DMP-} bond lengths arises from their the strong impact on the electron density of the PO$_2$ group and thus shielding properties of the $^{31}$P nuclear spin.
Short O1/O2-P and O3/O5-P bond lengths lead to an increase of electron density on the  PO$_2$ group, inducing a higher chemical shielding.
As the negative charge density is primarily located on the non-bridging O1/O2-P moiety, respective bond lengths show a stronger dependence than O3/O5-P bond lengths. In comparison, angular distortions of the phosphate tetrahedron and dihedral angle displacements have a minor effect on electron density distribution of the PO$_2$ group and thus change of shielding properties of $^{31}$P.\cite{Benda:JPCB:2012}

\begin{figure}
  \includegraphics[width=0.5\textwidth]{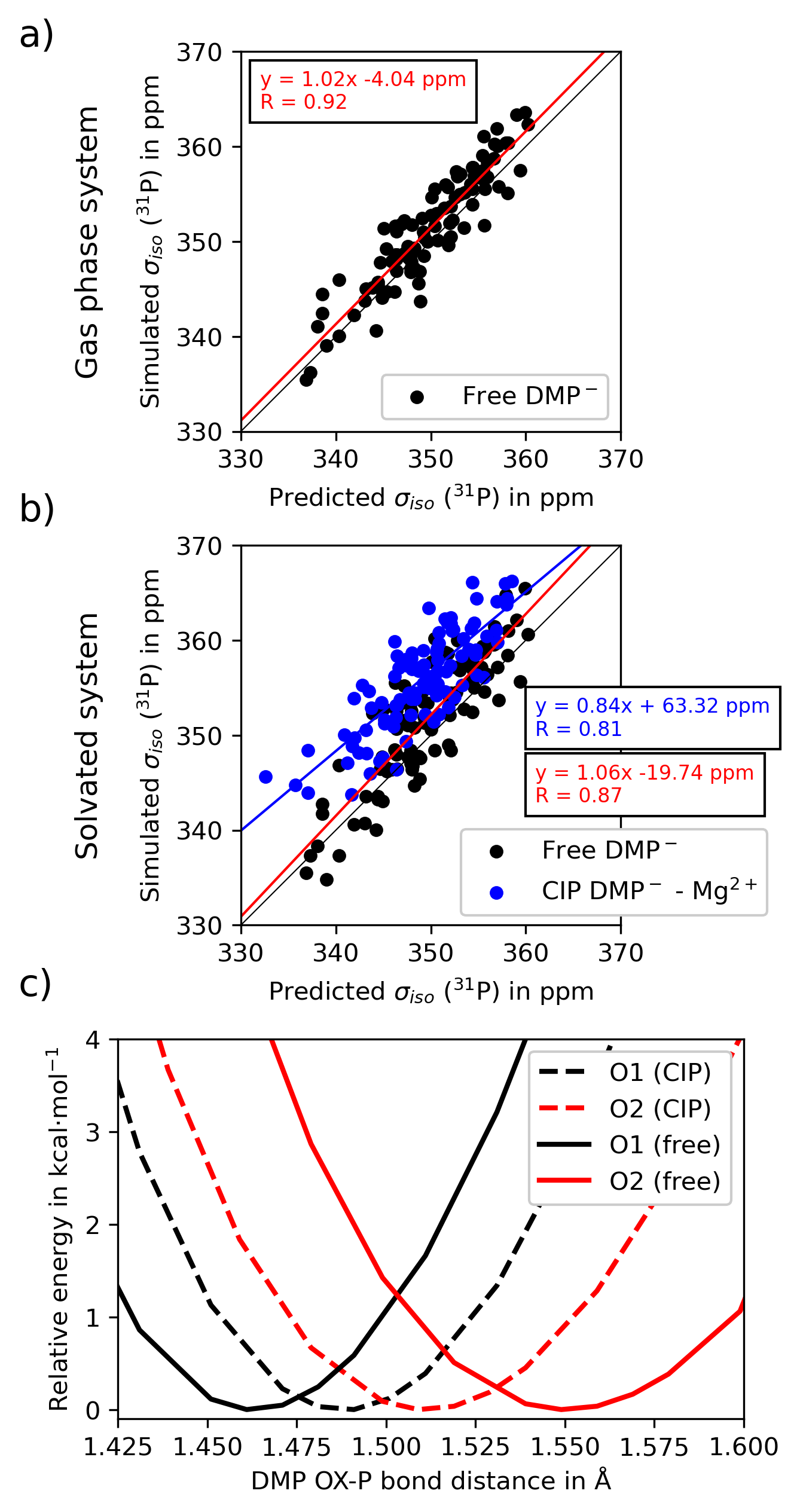}
  \caption{Geometric mapping of isotropic $^{31}$P shielding constants $\sigma_{iso}$($^{31}$P). (a) Correlation of ab initio simulated  and predicted $\sigma_{iso}$($^{31}$P) using  O1/O2-P and O3/O5-P bond lengths parameters for isolated \ce{DMP-} shown together with  the linear regression (red).
(b) Correlation of ab initio simulated  and predicted $\sigma_{iso}$($^{31}$P)  for free \ce{DMP-} in a 4.5~\AA ~solvation shell and  \ce{DMP-} - \ce{Mg^2+} CIP structures  using  O1/O2-P and O3/O5-P bond lengths parameters. Red - linear regression of \ce{DMP-} data points and blue - linear regression of CIP data points. (c) CIP-induced changes of bond lengths in hydration geometries of a \ce{DMP-} - \ce{Mg^2+} CIP. Shown are scans of  the O1-P and O2-P bond lengths in  the \ce{DMP-} - \ce{Mg^2+} CIP structure (dashed) and the equivalent \ce{DMP-} solvation structure without the \ce{Mg^2+} ion, evaluating the direct impact of the ion on bond length only.
}
  \label{fgr:figure_4}
\end{figure}

\paragraph{Geometric mapping of isotropic $^{31}$P shielding constants $\sigma_{iso}$($^{31}$P)}
Utilizing the strong geometric dependence, we aimed to establish a predictive mapping of ab initio simulated $\sigma_{iso}$($^{31}$P) from geometrical parameters.
For this purpose we evaluated the instantaneous O1/O2-P and O3/O5-P bond lengths of \ce{DMP-} in MD simulations and predict $\sigma_{iso}$($^{31}$P) from the 4D-coordinate map $\sigma_{iso}^{pred}(^{31}$P)($r_{O1-P}, r_{O2-P}, r_{O3-P}, r_{O5-P}$) (2D-surface maps are shown in Figs.~\ref{fgr:figure_3},c-d). 
Predictions 
are then initially compared to ab initio simulated $\sigma_{iso}$($^{31}$P) for the instantaneous  geometry of \ce{DMP-} in the gas phase.
Figure~\ref{fgr:figure_4},a shows the excellent correlation of this approach 
($R$ = 0.92, slope = 1.02), justifying the reduced dimensional mapping procedure from bond lengths and underscoring the minor role of  tetrahedral and dihedral angels of \ce{DMP-} for details of $\sigma_{iso}$($^{31}$P). Note that the error in $y$-crossing (-4.04 ppm) closely reflects the uncertainty introduced via the angular O-P-O coordinate.
%

In a second step, we partitioned the configurational ensemble sampled in MD simulations into \ce{DMP-} - \ce{Mg^2+} CIP structures and  free \ce{DMP-}.
For both sub-ensembles we evaluated $\sigma_{iso}$($^{31}$P) on the ab initio level of theory, accounting for the hydration shell up to 4.5~\AA, thus accounting for the polarization of \ce{DMP-} by the hydration shell (cf.~Fig.~\ref{fgr:figure_2},a) and interactions with the bivalent \ce{Mg^2+}  counter ion (cf.~Fig.~\ref{fgr:figure_3},a).
The ab initio simulated $\sigma_{iso}$($^{31}$P) of \ce{DMP-}$\times$ $N(\ce{H2O})$ w./w.o. \ce{Mg^2+} were then compared to the predictions from geometric bond-length parameters (Fig.~\ref{fgr:figure_4},b). 
Compared to the mapping of the constituent gas-phase geometries (Fig.~\ref{fgr:figure_4},a), the correlation for both sub-ensembles is reduced. 
Nevertheless, when comparing both sub-ensembles, it becomes apparent that for free \ce{DMP-} 
the correlation of predicted and ab initio simulated $^{31}$P $\sigma_{iso}$ is only slighly reduced ($R$ = 92 \emph{vs.} 87), also the slope of 1.06 is reasonable.
We interpret this finding that net influence of the full hydration environment of \ce{DMP-} on $\sigma_{iso}$($^{31}$P) is minor.
The spherical hydration shell to some extent polarizes the charge density of the PO$_2$ group but only has a  moderate impact on the 
shielding properties. 
Indeed, when comparing $\sigma_{iso}$($^{31}$P) simulated for \ce{DMP-} in water and  free \ce{DMP-} in vacuum, respective isotropic \ce{^31P} shielding constants only differ by 0.7 ppm (cf. Table~\ref{tbl:dmp_shielding}). For comparison, the geometric mapping prediction yields an error of 1.4 ppm. 

Comparing the CIP and the free \ce{DMP-} 
sub-ensembles, we find that for the former correlation coefficients are further reduced ($R$ = 81 \emph{vs.} 87) and linear regression yields a largely modified slope of 0.84 with a substantially upshifted y-axis zero-crossing (63.3 ppm).
Thus, $\sigma_{iso}$($^{31}$P) of the CIP sub-ensemble statistically 
occur shielded
compared to the free \ce{DMP-} counterpart.
The statistical analysis demonstrates that the predominant interaction leading to a breakdown of the geometric correlation of $\sigma_{iso}$($^{31}$P) 
arises from ion pairing with \ce{Mg^2+}.
$^{31}$P NMR is thus suggested to be particularly sensitive to the formation of CIP.

The sensitivity towards CIP is further corroborated by comparing $\sigma_{iso}$($^{31}$P) simulated for free \ce{DMP-} and 
\ce{DMP-} - \ce{Mg^2+} CIP
in water and  and vacuum. Both show a
shift $\Delta\sigma_{iso}$($^{31}$P)
by -3.9 ppm and -3.5 ppm, respectively (cf. Table~\ref{tbl:dmp_shielding}). 
A similar effect is found for \ce{DMP-} - \ce{Ca^2+} CIP, where the 
shift of $\sigma_{iso}$($^{31}$P) in water is even more pronounced ($\Delta\sigma_{iso}$($^{31}$P) = -7.6~ppm, Table~\ref{tbl:dmp_shielding}). Part of the larger shift for \ce{Ca^{2+}} is ascribed to the solvation shell that is more pronounced as compared to \ce{Mg^{2+}} (blue line in Fig.~\ref{fgr:figure_2},a).
\begin{table}
  \caption{Simulated ($\sigma_{iso}^{sim}$($^{31}$P)) and predicted ($\sigma_{iso}^{pred}$($^{31}$P)) isotropic \ce{^31P} shielding constants in \ce{DMP-}: 
   ab initio simulated $\sigma_{iso}^{sim}$($^{31}$P) are calculated on the GIAO-DF-LMP2 level of theory with a full pCVTZ basis set and a solvation shell of 4.5~\AA.
  }
  \label{tbl:dmp_shielding}
  \begin{tabular}{lll}
    \hline
     System & $\sigma_{iso}^{sim}$($^{31}$P) / ppm& $\sigma_{iso}^{pred}$($^{31}$P)  / ppm\\
    \hline
    Free \ce{DMP-} in water & 351.9 & - \\
    Free \ce{DMP-} in vacuum & 351.2  & 349.8\\
    & & \\

    CIP \ce{DMP-}-\ce{Mg^2+} in water  & 355.8  & - \\
    CIP \ce{DMP-}-\ce{Mg^2+} in vacuum  & 354.7  & 352.8 \\
    & & \\
    CIP \ce{DMP-}-\ce{Ca^2+} in water   & 359.5  & - \\
    CIP \ce{DMP-}-\ce{Ca^2+} in vacuum   & 354.3  & 357.9 \\
    \hline
  \end{tabular}
\end{table}

We ascribe the width of  the $\sigma_{iso}$($^{31}$P) distributions and the shift of mean along the 1D-reaction coordinate of CIP \ce{DMP-} - \ce{Mg^2+} CIP formation (cf.~Fig.~\ref{fgr:figure_3},a)
to two main structural factors, (i) the stochastic sampling of O-P bond lengths and (ii) CIP formation.
The simulations show that a Gaussian inhomogeneity on the order of \SIrange[range-phrase=\,--\,, range-units=single]{10}{20}{ppm} arises from the sampling of instantaneous \ce{DMP-} geometries, dominated by the distribution of P-O bond lengths. The width of the distributions is largely independent of details of the reaction coordinate of CIP formation. 
In contrast the increased shielding, reflected in the  
mean of simulated $\sigma_{iso}$($^{31}$P) for short ($\lt$ 2.5~\AA) \ce{DMP-} - \ce{Mg^2+} distances arises from the direct interaction with \ce{Mg^2+} ions upon CIP formation. Structurally, CIP formation is related to the replacement of a \ce{H2O} molecule in the octahedral first solvation shell  of the ion by one of the non-bridging oxygen atoms O1/O2 of \ce{DMP-} (cf.~Fig.~\ref{fgr:figure_1},a).


As $\sigma_{iso}$($^{31}$P) is highly sensitive to the local electronic environment and geometry of phosphorus atoms, 
the increased shielding of the $^{31}$P spin in CIP is 
potentially induced by the geometric details, like bond length changes of the PO$_2$ group upon CIP formation, a change of electron density via polarization of the PO$_2$ group by the adjacent \ce{Mg^2+} 
and/or electron density changes due to charge transfer between \ce{DMP-} and \ce{Mg^2+}.
We aimed to disentangle the origin of the shielding 
in \ce{DMP-} - \ce{Mg^2+} CIP
by evaluating the CIP-induced changes of 
bond lengths in hydration geometries of the \ce{DMP-} - \ce{Mg^2+} CIP 
(Fig.\ref{fgr:figure_4},c). Further we monitored changes of net atomic charges of solvated \ce{DMP-} and solvated \ce{DMP-} - \ce{Mg^2+} CIP 
(Table~\ref{tab:partial_charges_mg_ca}).
Fig.\ref{fgr:figure_4},c shows the scan of  O1-P and O2-P bond length in  the \ce{DMP-} - \ce{Mg^2+} CIP structure with and without \ce{Mg^2+}, preserving the hydration geometry and only evaluating the direct impact of the ion.
Due to the asymmetric  hydration environment in neither geometry O1-P and O2-P bond lengths are equivalent. Nevertheless, bond length changes upon addition or removal of the \ce{Mg^2+} ion are imprinted in mirror direction on the PO$_2$, i.e., a CIP-induced contraction of O2-P is accompanied by an elongation of O1-P by nearly the same amount. Thus,  the mapping of changes of O1-P and O2-P bond lengths  to 
changes of $\sigma_{iso}$($^{31}$P) (Fig.\ref{fgr:figure_3},c) suggests that the effects on $\sigma_{iso}$($^{31}$P) are nearly cancelled.

To the contrary, the analysis of atomic partial charge differences $\Delta q$ demonstrates the re-polarization and charge transfer upon CIP formation.
On average, the high charge density of  \ce{Mg^2+} leads to a charge transfer on the order of 0.13 $e^-$ between \ce{DMP-} and \ce{Mg^2+}
 (\ce{DMP^{0.87-}} - \ce{Mg^{1.87+}}).
The non-bridging oxygen O1 in contact  with \ce{Mg^2+} acquires substantial positive polarization 
($\Delta q_{O1} = 0.138 e^-$) 
that, besides charge transfer with \ce{Mg^2+}, is compensated by minor charge differences on bridging O3/O5 oxygen atoms and methyl groups. Thus, a net flux of charge from the non-bridging backbone region to the O1-P bond increases the charge density on the phosphorus atom ($\Delta q_{P} = -0.014 e^-$), correlating with increased shielding, while non-bridging O1 and O2 oxygen atoms acquire substantial charge asymmetry ($\Delta q_{O2} = -0.0004e^-$). 
Notably, charge density changes on the phosphorus atom are more pronounced for Ca$^{2+}$($\Delta q_{P} = -0.18 e^-$).
In summary our results suggest that the predominant interaction for the increased shielding of $\sigma_{iso}$($^{31}$P) is charge transfer, accompanied by a re-polarization of the PO$_2$ group, inducing a  net increase of electron density at the position of the phosphorus atom,
while geometric effects on $\sigma_{iso}$($^{31}$P) are minor.

\subsection{Quantitative analysis of contact ion pair formation}

Comparing the experimentally observed $\Delta\delta_{iso}$($^{31}$P) on order of \SI{-2}{ppm}
with simulated chemical shift differences $\Delta\delta^{sim}$($^{31}$P) 
of free \ce{DMP} and   \ce{DMP-} - \ce{Mg^2+} CIP 
\begin{equation}
\begin{aligned}
\Delta\delta^{sim}(^{31}\mathrm{P}) &= \sigma_{iso}^{Ref}(^{31}\mathrm{P}) - \sigma_{iso}(^{31}\mathrm{P})[\mathrm{CIP}]  -  (\sigma_{iso}^{Ref}(^{31}\mathrm{P}) - \sigma_{iso}(^{31}\mathrm{P})[\mathrm{Free} \ce{DMP-}])
\\
&= \sigma_{iso}(^{31}\mathrm{P})[\mathrm{Free} \ce{DMP-}] - \sigma_{iso}(^{31}\mathrm{P})[\mathrm{CIP}], 
\end{aligned}
\end{equation}
translates into $\Delta\delta^{sim}(^{31}\mathrm{P})$  with same sign as in the experiment, corresponding to a shielding of the $^{31}$P nucleus  in CIP (Table~\ref{tbl:dmp_shielding}). 
Nevertheless, a substantial discrepancy persists when comparing the amplitude of the chemical shift differences    $\Delta\delta_{iso}$($^{31}$P) and $\Delta\delta^{sim}(^{31}\mathrm{P})$, that is  on the order of  $\Delta\delta^{sim}(^{31}\mathrm{P})$  $\approx$ \SI{-3.9}{ppm}  in simulations.

Assuming a two component model, where deviations $\Delta\delta_{iso}$($^{31}$P) 
relative to free \ce{DMP-} (0~eq. \ce{Mg^2+}, cf. Fig~\ref{fgr:figure_1}) are induced by  \ce{DMP-} - \ce{Mg^{2+}} CIP, i.e., subsuming the population of fully solvated species SSIP with free \ce{DMP-}, 
the observed $\Delta\delta_{iso}$($^{31}$P) 
is only influenced by the relative population c[CIP] induced by the bimolecular formation of the \ce{DMP-} - \ce{Mg^2+} CIP  species:
\begin{equation}
\Delta\delta_{iso}(^{31}\mathrm{P}) \propto c[CIP], c[\mathrm{Free} \ce{DMP-}].
\end{equation}
Taking now the chemical shift values from simulations to characterize $\Delta\delta_{iso}^{max}$($^{31}$P) = $\Delta\delta^{sim}(^{31}\mathrm{P})$,
the maximal amplitude  $\Delta\delta_{iso}^{max}$($^{31}$P)= -3.9 ppm correlates  to a 
CIP formation of 100 \%.
The relative fractional population $X\left[ CIP\right] = n\left[ CIP\right] / n\left[ \mathrm{DMP}^{–}\right]$ 
can then be expressed as\cite{Haake:InorChem:84,Marr:JPCB:2024}
%
\begin{equation}
X[CIP] = \frac{\delta_{iso}(^{31}\mathrm{P})[x~\mathrm{eq.}~\ce{Mg^{2+}}] - \delta_{iso}(^{31}\mathrm{P})[0 ~\mathrm{eq.}~\ce{Mg^{2+}}]}{\Delta\delta_{iso}^{max}(^{31}\mathrm{P})}.
\label{eq:ModelCIP_2}
\end{equation}

\begin{figure}
  \centering
 \includegraphics[width=0.5\textwidth]{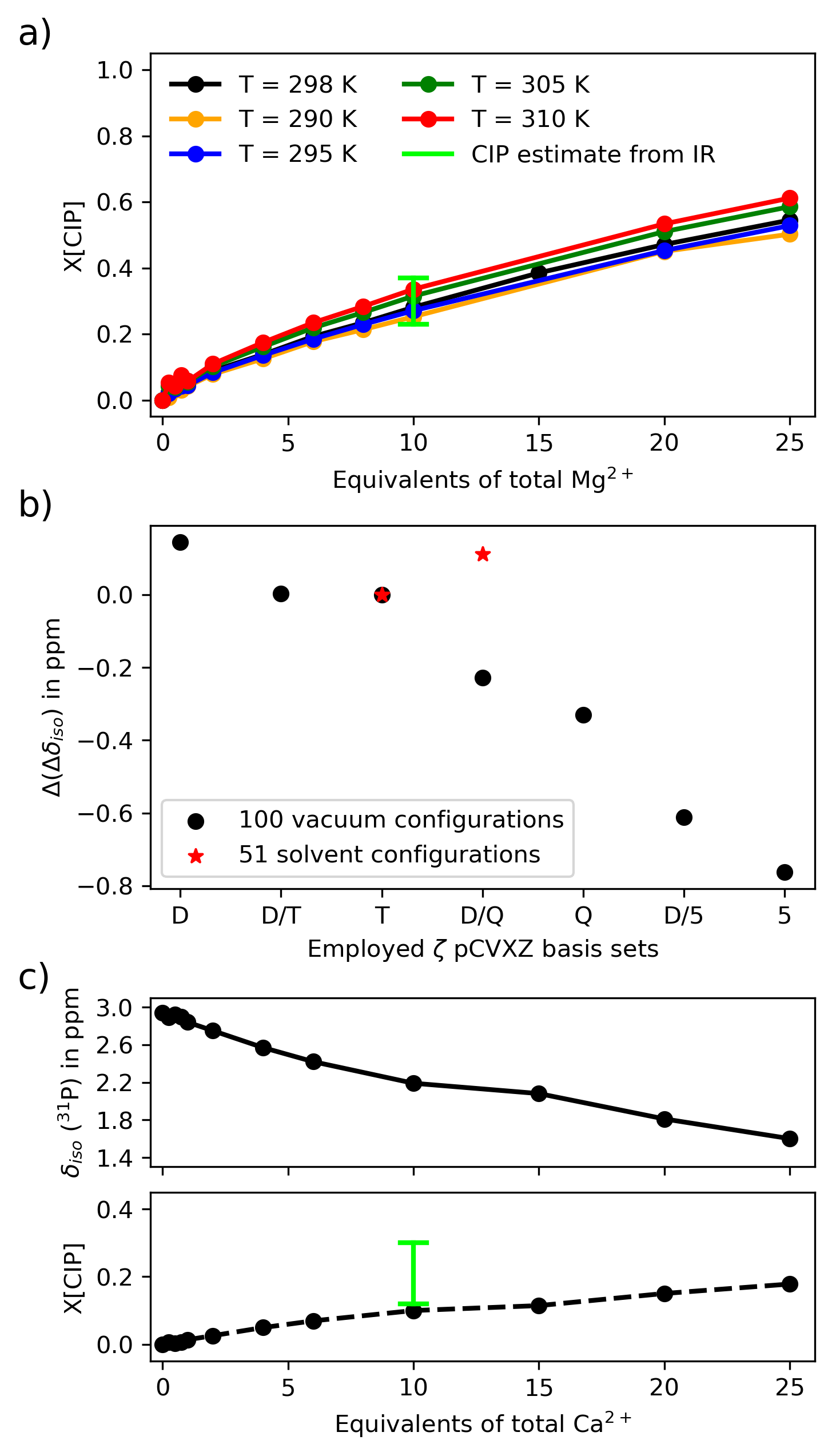}
  \caption{(a) Relative CIP population $X[CIP]$ (Eq.~\ref{eq:ModelCIP_2}) for varying \ce{Mg^2+} content and temperature in comparison to results from infrared spectroscopy\cite{Schauss:JPCL:2019_2} (green bars). 
  (b) Basis set dependence of amplitude of simulated chemical shift differences $\Delta\delta_{iso}^{max}$($^{31}$P) of free \ce{DMP-} and  \ce{DMP-} - \ce{Mg^2+} CIP (gas phase simulations), red stars show $\Delta\delta_{iso}^{max}$($^{31}$P) of free \ce{DMP-} and  \ce{DMP-} - \ce{Mg^2+} CIP  with water environment obtained with a  pCVQZ basis set.
  %
  (c) Observed chemical shifts $\delta(^{31}\mathrm{P})$ for varying \ce{Ca^2+} content (298 K) and derived \ce{DMP-} - \ce{Ca^{2+} }CIP population $X[CIP]$ (dashed line, lower panel). For comparison results derived from infrared spectroscopy are shown (green bars).
  %
  %
  %
}
  \label{fgr:figure_5}
\end{figure}

Figure~\ref{fgr:figure_5},a presents the relative fractional population $X[CIP]$, derived from eq.~\ref{eq:ModelCIP_2}, as a function of \ce{Mg{^{2+}}} content and temperature. The model suggests a variation of $X[CIP]$ in the range 
\SIrange[range-phrase=\,--\,, range-units=single]{0}{60}{\%} for up to 25 eq. \ce{Mg{^{2+}}}, i.e., saturation limit is not achieved in the investigated ion concentration range. 
Temperature variation in the range \SIrange[range-phrase=\,--\,, range-units=single]{17}{35}{\degreeCelsius} increases $X[CIP]$ by about \SI{10}{\%} for 10 eq. \ce{Mg{^{2+}}} and  by about \SI{15}{\%} for 25 eq. \ce{Mg{^{2+}}}.  
Comparing the $^{31}$P-NMR results for 10 eq. \ce{Mg{^{2+}}} (298~K,
$X[CIP]$ = \SI{28}{\%}) with values derived from infrared and 2D-infrared spectroscopy\cite{Schauss:JPCL:2019_2} (black line and green bars in Fig.~\ref{fgr:figure_5},a), 
the relative populations are in good agreement.

The determination of the relative CIP population $X[CIP]$  depends on the maximal amplitude from ab initio simulations  $\Delta\delta_{iso}^{max}$($^{31}$P) = $\Delta\delta^{sim}(^{31}\mathrm{P})$. To explore the sensitivity of our results, we investigated the dependence of $\Delta\delta^{sim}(^{31}\mathrm{P})$ on basis set size. As $\Delta\delta^{sim}(^{31}\mathrm{P})$ are very similar for \ce{DMP-} and  \ce{DMP-} - \ce{Mg^2+} CIP in the gas phase and in water ( \SI{-3.9}{ppm} and \SI{-3.5}{ppm}, respectively; Table~\ref{tbl:dmp_shielding}), we explored the convergence of $\Delta\delta^{sim}(^{31}\mathrm{P})$ up to quintuple zeta basis sets in gas phase simulations (Fig.~\ref{fgr:figure_5},b). We find that convergence for the ionic species is slow and deviations are on the order of \SIrange[range-phrase=\,--\,, range-units=single]{0.6}{0.8}{ppm}, corresponding to a \SIrange[range-phrase=\,--\,, range-units=single]{17.1}{22.9}{\%} uncertainty in the relative population  $X[CIP]$. 
For  aqueous  \ce{DMP-} and  \ce{DMP-} - \ce{Mg^2+} CIP  we find a reduced sensitivity on basis set (up to quadruple zeta), suggesting  a faster convergence of $\Delta\delta^{sim}(^{31}\mathrm{P})$ within \SI{0.2}{ppm} (red star in Fig.~\ref{fgr:figure_5},b; corresponding uncertainty in $X[CIP]$: \SIrange[range-phrase=\,--\,, range-units=single]{5}{10}{\%} )].

For comparison, Fig.~\ref{fgr:figure_5},c depicts 
$\delta_{iso}$($^{31}$P) for varying \ce{Ca^{2+}} concentrations, for which $\Delta\delta_{iso}$($^{31}$P) spans  a range of \SI{-1.4}{ppm}.
Combining these values 
with the chemical shift amplitude $\Delta\delta^{sim}(^{31}\mathrm{P})$ = \SI{-7.6}{ppm} 
in \ce{DMP-} - \ce{Ca^2+} CIP (see Table~\ref{tbl:dmp_shielding}) suggests
a reduced population $X[CIP]$ for \ce{Ca^{2+}} compared to \ce{Mg^{2+}}.
For 10 eq. \ce{Ca^{2+}} the derived population $X[CIP]$ = 0.1 is sightly outside the \SIrange[range-phrase=\,--\,, range-units=single]{12}{30}{\%} range suggested by infrared and 2D-infrared spectroscopy\cite{Schauss:JPCL:2019_2} (green bars in Fig.~\ref{fgr:figure_5},c).
Again, NMR derived concentrations more closely reflect the lower bound derived from infrared experiments.

%

\section{Discussion}

Experimentally, the increase in concentration of the bivalent counterion \ce{Mg^{2+}} in a range [0-25]~eq.  leads to a chnage of chemical shift 
$\Delta\delta_{iso}$($^{31}$P) = \SI{-2.07}{ppm} in \ce{DMP-},
i.e., a shielding of the $^{31}$P nucleus with moderate temperature dependence. 
The magnitude of  $\Delta\delta_{iso}$($^{31}$P) is comparable but reduced (\SI{-1.4}{ppm}) for \ce{Ca^{2+}} at room temperature.
Notably, for both bivalent counterions, saturation behavior is not observed, even for the highest investigated ion content. 
High level ab initio theory (GIAO-DF-LMP2) predicts a shielding of the $^{31}$P spin in 
CIP by $\Delta\delta^{sim}$($^{31}$P) = \SI{-3.9}{ppm} and \SI{-7.6}{ppm} for \ce{Mg^{2+}} and \ce{Ca^{2+}}, respectively, that is comparable in magnitude to studies on DNA - \ce{Mg^{2+}} binding using density functional theory.\cite{Benda:JPCA:2011}
%
While both theory and experiments show a shielding of the $^{31}$P spin, 
the simulations for both ions 
suggest a substantially larger chemical shift amplitude $\Delta\delta^{sim}$($^{31}$P) associated with CIP formation.

 The observations were analyzed in a two-component model (Eq.~\ref{eq:ModelCIP_2}), where $\delta_{iso}$($^{31}$P) is determined by the average chemical shift of \ce{DMP-} and $X[CIP]$, the unknown relative fraction of \ce{DMP-} - \ce{Mg^2+} CIP.
As such, the model assumes the fast exchange of solvation species on the timescale of the experiment, leading to a single average peak $\delta^{sim}$($^{31}$P) in the $^{31}$P-NMR spectra (Fig~\ref{fgr:figure_1},c).
Further, fully solvated \ce{DMP-} species, i.e. SSIP and free \ce{DMP-}, are assumed to be indistinguishable in the $^{31}$P-NMR experiment, 
which is motivated by the minor chemical shift differences of SSIP and free  \ce{DMP-} in simulations (Fig.~\ref{fgr:figure_3},a). 
The model reproduces two key observations: (i) the missing saturation behavior in the investigated concentration range and (ii) the moderate temperature dependence of $\delta_{iso}$($^{31}$P). 
%
The relative CIP fraction $X[CIP]$ was then determined in a broad range of \ce{Mg^{2+}} concentrations up to 25 eq., approaching $\approx$ 50 $\pm$ 10 \% at room temperature, where the uncertainty is estimated from the variation of the chemical shift amplitude $\Delta\delta^{sim}$($^{31}$P) (Fig.~\ref{fgr:figure_5},b).


While similar two-component models have been invoked before to describe ion induced chemical shift changes in $^{31}$P NMR spectra,\cite{Kutus:JMolLiqu:22,Haake:InorChem:84} 
the assignment 
so far assumed an unspecified cationic bound species, tentatively assigned to ion pairing, 
without molecular distinction of the strongly interacting CIP and SSIP  species.
In the current work, we provide an in-depth ab initio analysis of the microscopic mechanism leading to the observed chemical shift changes associated with the molecular solvation species (CIP, SSIP and free \ce{DMP-}, respectively). Further the influence of the solution environment is explored in comparison to  gas phase simulations.
The ab initio determination of  $\sigma^{iso}$($^{31}$P) 
strongly suggests that the $^{31}$P chemical shifts of \ce{DMP-} are particularly sensitive to the formation of CIP (Fig.~\ref{fgr:figure_3},a).
Moreover, the explored concentration range is much broader than in previously reported $^{31}$P-NMR experiments.
Association constants derived from dielectric relaxation and $^{31}$P-NMR spectroscopy agree well at low concentrations ($<$ 2 eq.). For multiple eq. of \ce{Mg^2+} or \ce{Ca^2+} ($>$4) the  ion populations derived from a fit of the amplitude $\Delta \delta_{iso}^{max}(^{31}$P) acquire increasing  uncertainty, presumable due to secondary ion hydration equilibria.\cite{Buchner:PCCP:2009}
Note that even for highest Mg$^{2+}$ content (25 eq.), the ion concentration is about 
2.2 M, i.e., enough water for the full solvation of ionic species is assured under the current experimental conditions.


The high sensitivity  of $^{31}$P NMR to detect CIP of \ce{DMP-} with bivalent ions \ce{Mg^2+} and \ce{Ca^2+} resembles findings from IR and 2D-IR spectroscopy that selectively report on contact ion pair formation with PO$_2$ groups.\cite{Schauss:JPCL:2019,Schauss:JPCL:2019_2,Schauss:JPCB:2021} 
The derived relative CIP fractions $X[CIP]$  at 10 eq. of ions 
agree reasonably (Figs.~\ref{fgr:figure_5},a,c). 
Despite these comparable analytical findings, the underlying mechanism responsible for high species sensitivity is vastly different in the two experimental techniques. 
The short P-O$^-$....\ce{Mg^{2+}} distance in CIP of about \SI{2.1}{\angstrom} induces
a strong perturbation of the PO$_2$ mode potential, primarily arising from exchange repulsion interactions of the closed shell \ce{Mg^{2+}} and PO$_2$ electron density.
In the IR-spectra, this leads to a characteristic blue shift on the order of 20-30 cm$^{-1}$ of the asymmetric PO$_2$ stretching vibration, compared to fully solvated PO$_2$ groups.\cite{Schauss:JPCL:2019,Schauss:JPCL:2019_2}
Considering the typical linewidths of the asymmetric PO$_2$ stretching vibration
(on the order of \SIrange[range-phrase=\,--\,, range-units=single]{30}{70}{\per\centi\meter} in DNA and RNA\cite{Kundu:JPCB:2020}), the CIP-induced blue shift 
leads to a shoulder in the IR spectra of the asymmetric PO$_2$ stretching vibration that can be accentuated via non-linear 2D-techniques. 
The lineshapes in 2D-IR spectra are characterized to a large extent by a quasi-homogeneous contribution from fast (sub-\SI{100}{\femto\second}) structural fluctuations\cite{Elsaesser:JPCB:2021} and, on the timescale of the IR experiments (femto- to picoseconds), longer-lived structural inhomogeneities.
The IR lineshapes thus mirror  
the population of all species and encode details of the CIP population and the hydration environment.

In contrast, $^{31}$P-NMR spectra report on 
changes of 
electron density of the PO$_2$ group upon CIP formation. 
The observation of
the shielding of the $^{31}$P in CIP hydration structures
is a signature of the modified charge density, where the strong interaction with \ce{Mg^{2+}} and \ce{Ca^{2+}} leads to charge transfer on the order of 0.13 $e^-$ between CIP constituents and a  polarization of the PO$_2$ group that breaks the symmetry between the two non-bridging oxygen atoms.
The 
$^{31}$P relaxation rates reflect the fast motional regime (extreme narrowing limit, $\omega_0 \tau_L << 1$) and in addition a contribution from 
%
molecular exchange events between CIP  and SSIP or free \ce{DMP-}.
The population weighted 
average chemical shift $\delta_{iso}$($^{31}$P) of CIP and free \ce{DMP-} is observed for every counter ion concentration (cf. Fig.~\ref{fgr:figure_1},c).   
The magnitude of $\Delta\delta_{iso}$($^{31}$P) ($\approx$ \SI{-2}{ppm}) is substantially larger than the NMR linewidths. 
Thus, $^{31}$P-NMR spectroscopy  provides in principle a higher sensitivity than IR spectroscopy at low bivalent ion concentrations.
The technique further offers potential 
to discriminate long- and short-lived ion pair species with PO$_2$ groups.
For example, $^{31}$P NMR spectra of tRNA \cite{Gueron:PNAS:75} show the emergence of individual lines for increasing \ce{Mg^{2+}} concentration. 
%
%
%
%
%
%
The observed exchange broadenings 
with increasing \ce{Mg^{2+}} concentration (Table~\ref{tbl:exp}) are 
well documented.\cite{Rao:JBioChem:76,Granot:Biopol:82}
Based on the calculated $\Delta\delta^{sim}(^{31}$P), the exchange rate for which coalescence of the resonances lines of free \ce{DMP-} and CIP is expected is about 2800 s$^{-1}$, corresponding to a lifetime of 350 $\mu$s of the CIP.
%
Estimates of the exchange rate have placed a lower limit on the exchange time scale of 1 $\mu$s\cite{Freisinger:CoordChemRev:2007}.

%

The theoretical assignment of the observed chemical shift changes $\Delta\delta_{iso}$($^{31}$P) 
to CIP formation is further corroborated by $^{31}$P-NMR experiments with \ce{Zn^{2+}}, \ce{Cd^{2+}}, \ce{Co(NH_3)6^{3+}} and \ce{Na^{+}} counter ions (summarized in Table~\ref{tbl:exp}).
Among the latter, only 
\ce{Zn^{2+}} induces a comparable shift of $\delta_{iso}$($^{31}$P), suggesting the formation of a substantial fraction of CIP. Severe perturbation of the $^{31}$P-NMR lineshape precludes quantitative analysis for $>$ 0.1 eq. 
While for \ce{Na^+} the minor deshielding 
($\Delta\delta_{iso}$($^{31}$P) = \SI{0.07}{ppm} )
is well documented and correlates with the minor tendency of \ce{DMP^-} - \ce{Na^+} CIP formation suggested by IR spectroscopy\cite{Schauss:JPCL:2019_2}, 
the absence of a substantial $\Delta\delta_{iso}$($^{31}$P)  for  \ce{Co(NH_3)6^{3+}} strongly supports the notion that observed $^{31}$P chemical shift changes are induced by CIP formation.
Despite the high charge density of \ce{Co^3+},  \ce{Co(NH_3)6^{3+}} is well established to interact via outer sphere contacts only\cite{Gong:Biochem:2009}
and the $^{31}$P experiments for relatively high \ce{Co(NH_3)6^{3+}}  content (4-10 eq.) demonstrate that outer sphere interaction of \ce{DMP^-} with multivalent ions has negligible effect on  $\delta_{iso}$($^{31}$P). 
%
\emph{Vice versa}, the minor  $\Delta\delta_{iso}$($^{31}$P)  observed for \ce{Cd^{2+}} is in line with explorative MD that show a  spontaneous transformation of CIP into SSIP and free \ce{DMP-} on a $<$ 100 ns timescale, 
suggesting a minor role of CIP in solutions of \ce{DMP-} and \ce{Cd^{2+}}. 
%
%
NMR 
should thus be considered an effective short range reporter with particular CIP sensitivity, providing a local view of the electrostatics of PO$_2$ groups. 

In the simulations, the width of the $\sigma_{iso}$($^{31}$P) distributions (standard deviation $\approx$ 12 ppm) is much broader
than the linewidth of the $^{31}$P-NMR spectra and largely independent of the \ce{DMP^-} - ion separation (Fig.~\ref{fgr:figure_3}).
As \ce{DMP^-} - \ce{Mg^2+} geometries are sampled from MD simulations,  the statistical sampling of P-O bond lengths is directly reflected 
in the width of simulated $\delta$($^{31}$P) distributions.
Due to the particular short-range sensitivity of $\sigma_{iso}$($^{31}$P) to variations of the local charge density, the statistical sampling of P-O bond lengths was shown to be the dominant contribution to the distributional width.
Exploiting this relation allowed us to develop a low-dimensional geometric mapping procedure of $\sigma_{iso}$($^{31}$P) of \ce{DMP^-} that utilizes the instantaneous P-O bond lengths of the MD trajectory (Fig.~\ref{fgr:figure_4}).
The mapping approach provides quantitative insight into the distributional widths, allows to  de-noise the real-time data of MD simulations for the determination of $\delta_{iso}$($^{31}$P) in close comparison to NMR experiments  
and suggests that 
contributions from O-P-O angles and backbone conformation are minor.\cite{Benda:JPCB:2012}
Moreover the mapping approach establishes a statistical foundation for $\Delta\delta_{iso}$($^{31}$P) induced by CIP geometries that is encoded in the shift of the mean the distributions of $\delta_{iso}$($^{31}$P).
Similar strong dependence of $\delta$($^{31}$P) on P-O bond lengths was previously reported from  NMR spectroscopy of crystalline samples.\cite{Cheetham_JChemSoc_86} 

\begin{figure}
 \includegraphics[width=0.5\textwidth]{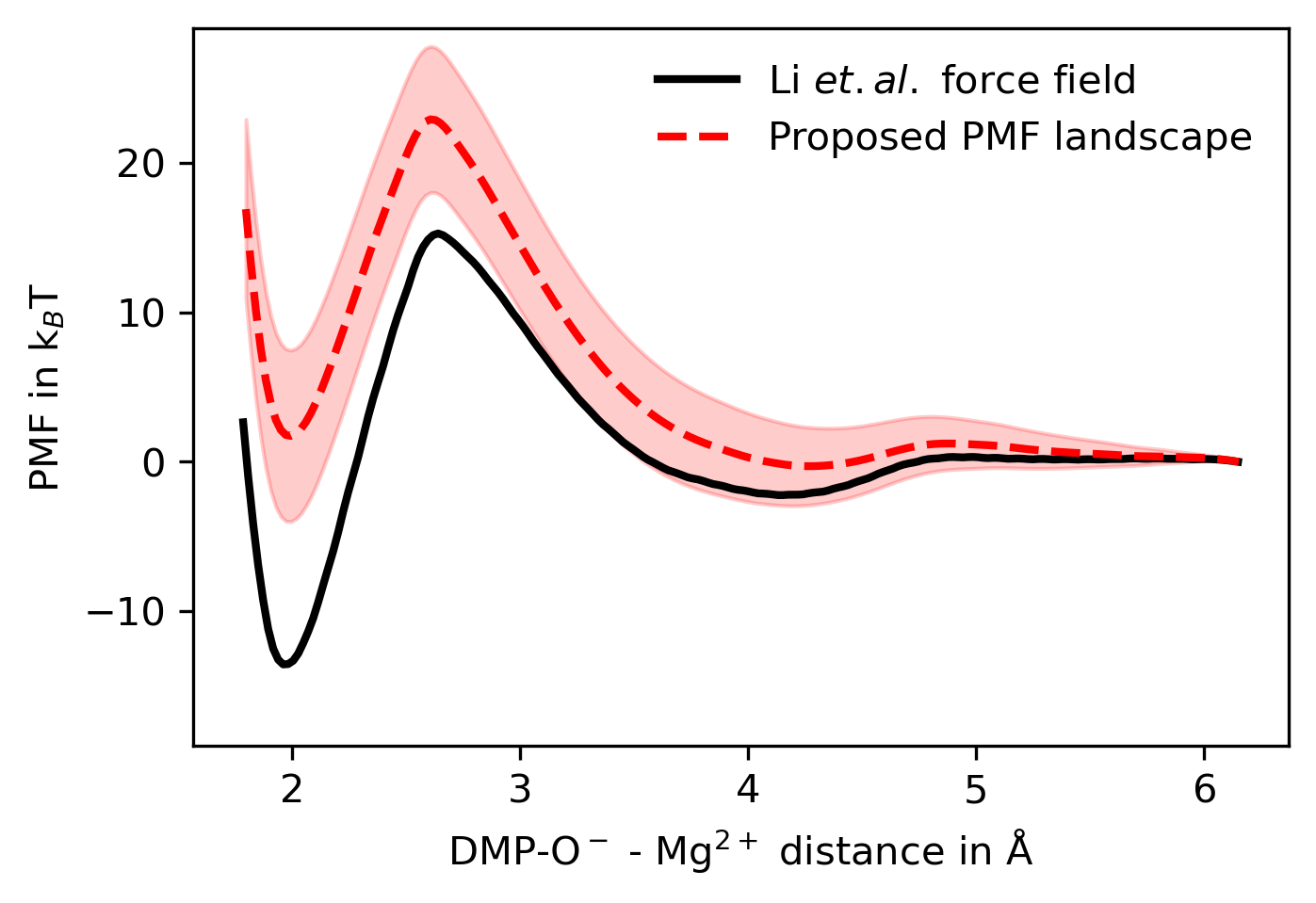}
  \caption{Relative stability of CIP and SSIP in solution. Minima of the CIP and SSIP species in the potential of mean force (PMF) determined by the 12-6-4 parametrization of intermolecular interactions  by Li \emph{et al.}\cite{Li:JCTC:2013,li_systematic_2020} (black) and from boundary conditions imposed on the relative population $X[CIP]$ derived in the current study (red dashed).
}
  \label{fgr:si_1}
\end{figure}
Utilizing $\Delta\delta_{iso}^{max}$($^{31}$P) = $\Delta\delta^{sim}$($^{31}$P) in a two-component analysis of contact and fully solvated \ce{DMP^-} species, allows to establish the relative population $X[CIP]$ of contact ion pairs, that form a minor species at low ion content ($<$ 10 eq.) and room temperature ($X[CIP]$ = \SI{28}{\%}). 
Knowledge of the  relative population $X[CIP]$ places boundary conditions on the relative stability of CIP and SSIP in solution (Fig.~\ref{fgr:si_1}).
Specifically, 
the free energy of SSIP or free \ce{DMP-} solvation species are required to be more stable than the minimum of the CIP species in the potential of mean force (PMF) and the moderate temperature dependence establishes that both minima are energetically close within a few k$_B$T.
In contrast, common fixed-charge force fields\cite{Li:JCTC:2013,li_systematic_2020} yield CIP structures that are substantially more stable than SSIP and free \ce{DMP^-} (cf. Fig.~\ref{fgr:si_1}, black line), with the tendency of overbinding to the phosphate groups of DNA and RNA.
Recent investigations on \ce{DMP^-} - ion interactions\cite{Puyo-Fourtine:JPCB:2022} indicate that the original parametrization of the polarizable amoeba force filed overestimates \ce{Mg^{2+}} and \ce{Ca^{2+}} binding.
The results of Puyo-Fourtine \emph{et al.}\cite{Li:JCTC:2013,li_systematic_2020} further suggest that force fields with overall reasonable ion binding free energies tend to underestimate CIP population in comparison to 2D-IR experimental estimates, indicating an over-stabilization of SSIP.
%

Our results shed light on the relative driving force of bivalent ions \ce{Mg^{2+}} and \ce{Ca^{2+}} to form CIP with the PO$_2$ group.  Results from RISM theory suggest a preference of \ce{Ca^{2+}} to form CIP in complex RNA structures.\cite{Nguyen:JPCB:2020}
%
In the \ce{DMP^-} model  system, 
experimental and in-depth simulations of the $^{31}$P chemical shifts
allowed us to firmly establish that at given concentration \ce{Mg^{2+}} forms about twice as much CIP as \ce{Ca^{2+}}, in agreement with findings from IR spectroscopy.\cite{Schauss:JPCL:2019_2}
Concerning the primary PO$_2$ - ion interactions, hardness of the ion
and the properties of the \ce{I^{2+}(H_2O)_N} solvation shell are the decisive factors for CIP formation.
In particular, for \ce{Mg^{2+}(H_2O)_6}, the geometric constraints and desolvation of the first octahedral
solvation shell around the ion introduces an substantial energetic penalty that destabilizes the CIP species relative to the SSIP species.
Our results establish the microscopic origin of 
$^{31}$P chemical shifts induced by the bare electrostatic interaction of \ce{DMP-} and bivalent counter ions, establishing a reference to quantify
secondary interactions in the more complex hydration geometries of DNA and RNA.\cite{Kundu:JPCB:2020} 

\section{Conclusions}

Through the combination of NMR experiments and in-depth ab initio simulations of the $^{31}$P chemical shift of \ce{DMP^-} in aqueous solution, a model system of the DNA or RNA sugar-phosphate backbone, we have established the origin for the shielding of the $^{31}$P spin upon addition of the bivalent ions \ce{Mg^{2+}} and \ce{Ca^{2+}}. 
Specifically, electronic effects arising from the formation of \ce{DMP-} - \ce{I^{2+}} CIP structures (I = Mg, Ca) are distinguished from intramolecular geometric distortions of bridging and non-bridging P-O bond lengths of the phosphate group.
A low-dimensional geometric mapping procedure allows us to quantitatively describe the width of $\sigma_{iso}$($^{31}$P) distributions, establishing a \SI{-3.9}{ppm} amplitude of $^{31}$P chemical shift change ascribed to CIP formation with \ce{Mg^{2+}}. 
Such microscopic values explain the experimental $^{31}$P-NMR spectra in a vast concentration and temperature range. The data suggest that CIP are a minor species at low \ce{Mg^{2+}} concentrations and a relative CIP population $X[CIP]$ approaching \SI{50}{\%} only occurs for large \ce{Mg^{2+}} ion excess. Despite a larger $^{31}$P chemical shift amplitude (\SI{-7.6}{ppm}), \ce{Ca^{2+}} has a lower tendency of CIP formation. 
The derived CIP populations serve as benchmark for future force field optimization and refinement.

 \begin{acknowledgement}

This research has received funding from the Deutsche Forschungsgemeinschaft (DFG)  [SFB1309-C09, project number 325871075].
This work was funded by the DFG through the Cluster of Excellence $e$-conversion [Grant No. EXC2089/2-390776260, BPF] and  
the Emmy Noether program (project number 394455587 to AKS).
We thank Peter Dowling for synthesis of the hexaamminecobalt(III) chloride.
The authors acknowledge the computational and data resources provided by the Leibniz Supercomputing Centre (www.lrz.de).



 \end{acknowledgement}




\bibliography{achemso-demo}

\end{document}

%% file: nmr_methods/nmr_methods_2.tex
\subsubsection{Reagent preparation}
Chemicals were bought from Grüssing (Filsum, Germany), Sigma-Aldrich (Darmstadt, Germany) and Santa Cruz Biotechnology (Dallas, USA) and used as received. \ce{D2O} (\SI{99.9}{\%} D) was purchased from Deutero (Kastellaun, Germany).
Hexaamminecobalt(III) chloride was synthesized according to a modified literature synthesis\cite{mcreynolds:inorsynt:1946} by oxidation of an aqueous solution of Cobalt(II) chloride with hydrogen peroxide in presence of ammonium chloride, activated charcoal and ammonia at \SI{0}{\degreeCelsius}. A pure product could be obtained by precipitation with hydrochloric acid, followed by filtration with hot hydrochloric acid and recrystallization in ethanol.

\subsubsection{NMR experiments}
All NMR spectra were acquired on a Bruker Avance III HD 400 MHz spectrometer (\SI{9.4}{\tesla}), using a broadband BBO 400S1 probe. The measurements were set up, performed and evaluated using Bruker TopSpin 3.5. Relaxation data were analyzed with Dynamics Center 2.1.8. The temperature was regulated using the spectrometer's internal temperature control, after calibration with methanol and ethylene glycol\cite{HOFFMAN200587,AMMANN1982319}. 
\ce{^1H} spectra were first recorded without water suppression 
to obtain the shift of the maximum water peak height. This value was used as the offset for a second \ce{^1H} spectrum to ensure the maximum water signal reduction via the WATERGATE pulse sequence.
The \ce{^1H} pulse length was optimized after every titration step.
\ce{^13C} Spectra were obtained 
with a fixed \SI{30}{\degree} pulse. \ce{^31P} Spectra were acquired both with and without 
\ce{^1H} garp decoupling.
$T_1$ and $T_2$ relaxation times were measured for \ce{^31P} at \SI{25}{\degreeCelsius} in all ion titrations.
The inversion recovery method was used to determine $T_1$ relaxation times.
The Carr-Purcell-Meiboom-Gill (CPMG) pulse sequence was used to determine $T_2$ relaxation times 
using (\SI{20}{\milli\second} $\pi$ \SI{20}{\milli\second}) spin echo blocks. 
A monoexponential decay function was fitted to the signal intensities.
Chemical shifts were referenced against 3-(trimethylsilyl)-1-propanesulfonic acid sodium salt (DSS) for the initial spectra. Linewidths were automatically determined as full width at half height (FWHH) using the peakw command from TopSpin. For signals where the automatic FWHH determination failed due to broadness or overlap, the linewidths were determined manually.

\subsubsection{Titration experiments}
Samples of sodium dimethyl phosphate (\ce{NaDMP}) for titrations were prepared by adding \SI{450}{\micro\liter} of a 9:1 \ce{H2O}:\ce{D2O} mixture to \ce{NaDMP} powder (\SI{0.09}{\milli\mol}, \SI{0.01332}{\gram}, \SI{0.2}{M}) in a \SI{5}{\milli\meter} NMR tube with an additional \SI{1}{\micro\liter} of 
\SI{1}{mM} DSS solution. Stocks solutions with dissolved metal ion salts were titrated to achieve the required concentrations. Where applicable, the samples were measured at each reported temperature before adding new stock solution. The respective molar ratios n(ion)/n(\ce{DMP^-}) 
were expressed as equivalents (eq.) of metal ions to \ce{DMP-}. Series from \SI{0.1}{eq.} up to \SI{25}{eq.} were measured, depending on the effect of the metal ions on the NMR spectra.
Stocks solutions of the used ions and respective  salts are summarized in Table~\ref{tbl:sample_prepartation}.

%% file: tc_methods/md_methods.tex
Atomistic MD Simulations were performed with the MPI implementation of the Amber18 and Amber24 software packages\cite{amber24:Case:JChemInfMod:2025} and employed the parmbsc1 DNA force field parameters~\cite{Ivani2016parmbsc1} and partial charges of the phosphate group for \ce{DMP-} together with the TIP4P-FB water model~\cite{wang_building_2014} and corresponding 12-6-4 Lennard-Jones parameters for mono and divalent ions.\cite{Li:JCTC:2013,li_systematic_2020,sengupta_parameterization_2021} 
Systems were created with the LEaP software included in AmberTools24.\cite{ambertools:Case:JChemInfMod:2023} Analysis and visualization of the trajectories were carried out with 
pytraj as Python package binding to the cpptraj program\cite{Roe:JCTC:2013} and VMD~\cite{Humphrey:JMolGraph:96}.

Solvatization of the systems was realized in a truncated octahedron box with a minimum buffer size of \SI{20}{\angstrom}. Sodium cations or chloride anions were added for charge neutralization.
A cut-off criterion of \SI{9}{\angstrom} was imposed on the non-bonded interactions and long-range Coulomb interactions were treated with the particle mesh Ewald summation method\cite{darden_particle_1993,essmann_smooth_1995}. A time step of \SI{2}{\femto\second} was used for the integration of motions while all bonds including hydrogen and all water molecules were constrained with the SHAKE\cite{ryckaert_numerical_1977} and SETTLE\cite{miyamoto_settle_1992} algorithms, respectively.
All systems were optimized in a three-step approach by first applying harmonic positional constraints of \SI{100}{\kcal\per\mol\per\square\angstrom} on the solute atoms to allow relaxation of the solvent (10000 steps) followed by a second optimization with reduced harmonic restraints of \SI{10}{\kcal\per\mol\per\square\angstrom} (10000 steps) and a third constraint free optimization (10000 steps). After half of the respective optimization steps, the algorithm was switched from the steepest descent to the conjugated gradient method. The systems were heated over \SI{0.2}{\nano\second} to the target temperature of \SI{298}{\kelvin} using the Langevin thermostat\cite{loncharich_langevin_1992} employing a collision frequency of \SI{1}{\per\pico\second} which was followed by \SI{0.6}{\nano\second} of pressure equilibration to \SI{1}{bar} with the Berendsen barostat\cite{berendsen_molecular_1984} using a pressure coupling constant of \SI{1}{\pico\second}. During this equilibration period, harmonic positional constraints of \SI{100}{\kcal\per\mol\per\square\angstrom} were applied to the solute atoms which were gradually removed over the following \SI{1.2}{\nano\second} resulting in a final \SI{1.2}{\nano\second} unrestrained equilibration in the NpT ensemble.
The unbiased \ce{DMP-}, \ce{DMP-} - \ce{Mg^2+} and \ce{DMP-} - \ce{Ca^2+} systems were sampled for \SI{10}{\nano\second}. 

Umbrella sampling was performed with the non-bridging \ce{DMP-}-O1 - \ce{Mg^2+} distance (force field atom type O1P, see Fig.~\ref{fgr:figure_1},a for labeling) as reaction coordinate along which the potential of mean force (PMF) was evaluated from \SIrange{2.00}{6.00}{\angstrom} ($\Delta d = 0.1$\,\si{\angstrom}). 
A distance based restraint of \SI{300}{\kcal\per\mol\per\square\angstrom} was imposed on a virtual O1 - \ce{Mg^2+} bond. 
Each distance was initially re-equilibrated for at least \SI{0.1}{\nano\second}, starting from the nearest available distance sampling point with a subsequent sampling period of \SI{10}{\nano\second}. 
To avoid unwanted interactions between \ce{Mg^2+} and the second non-bridging oxygen atom O2 of \ce{DMP-} (force field atom type O2P), a repulsive harmonic potential between \ce{DMP-}-O2 - \ce{Mg^2+} with a force constant of \SI{30}{\kcal\per\mol\per\square\angstrom} was applied if the respective distance fell below \SI{3.4}{\angstrom}. Distance information was saved every \SI{0.1}{\pico\second}. The unbiasing of the umbrella sampling runs was performed with the vFEP\cite{vfep1,vfep2} routine with Jacobian corrections to obtain the free energy landscape.

Ab initio simulations of NMR shielding constants (see below, Fig.~\ref{fgr:figure_3},a) were performed with a sampling of linear P-O1-\ce{Mg^2+} configurations. Starting structures were taken from the umbrella sampling runs and \SI{600}{\kcal\per\mol\per\square\angstrom} angular and distance based harmonic restraints were subsequently applied to the P-O1-\ce{Mg^2+} atoms. The systems were re-equilibrated for \SI{0.4}{\nano\second}, followed by \SI{10}{\nano\second} production sampling runs, coordinates were saved every \SI{20}{\pico\second}.


%% file: tc_methods/qm_methods.tex

Nuclear shielding constants $\sigma_{iso}(^{31}$P) were calculated on the GIAO-DF-LMP2 ab initio level of theory\cite{Loibl:MolPhys:2010,Loibl:JCP:2012} implemented in Molpro 2024.1.\cite{Werner:JCP:2020,Werner:WIREComMolSci:2012} 
GIAO-MP2  simulations generally provide accurate chemical shifts (comparable or superior to DFT) over a broad range (\(\sim 600\) ppm) of phosphorus shielding constants,\cite{Rusakova:IJMS:2026} particularly when used with large basis sets and the GIAO-DF-LMP2 approximation retains very high accuracy.\cite{Loibl:JCP:2012}
In benchmark simulations we tested the quality of the basis set for $\sigma_{iso}(^{31}$P) and investigated the influence of the size of hydration shell. Benchmark simulations were performed on a randomly chosen configuration of a  \ce{DMP-} - \ce{Mg^{2+}} CIP and free   \ce{DMP-}
from MD simulations. 
Correlation-consistent polarized Dunning basis sets with core correlation (cc-pCVXZ, X=D,T,Q,5) were used for heavy atoms, while for hydrogen atoms cc-pVXZ (X=D,T,Q,5) were used. 
For density fitting\cite{Werner:JCP:2003} the corresponding basis sets without core correlation were applied.
Benchmarks have shown an insignificant contribution of the density fitting basis set on the nuclear shielding constants. 
As simulations on the  GIAO-DF-LMP2 level of theory are computationally expensive, in particular for system sizes with large hydratio shell (see below),
we additionally benchmarked mixed basis sets where hydrogen atoms were treated with double zeta accuracy (cc-pVDZ). 

The influence of an explicit solvent shell on $\sigma_{iso}(^{31}$P) of \ce{DMP-} was benchmarked for \ce{DMP-} - \ce{Mg^{2+}} CIP, \ce{DMP-} - \ce{Ca^{2+}} CIP and  free \ce{DMP-}.
The hydration shells were constructed with a distance based cut-off criterion from the surface of the solute molecules (\ce{DMP-} - \ce{Mg^{2+}} CIP, \ce{DMP-} - \ce{Ca^{2+}} CIP or \ce{DMP-}).
Nuclear shielding constants were averaged over 100 configurations taken from MD simulations with a relative spacing of \SI{0.1}{\nano\second} to assure uncorrelated configurations. Due to convergence issues four \ce{DMP-} - \ce{Ca^{2+}} configurations were discarded. 
To quantify the influence of the hydration shell on $\sigma_{iso}(^{31}$P), respective vacuum systems of \ce{DMP-} - \ce{Mg^{2+}} CIP, \ce{DMP-} - \ce{Ca^{2+}} CIP and  \ce{DMP-} were constructed by removing the solvent molecules.

Atomic partial charges were evaluated using the Mulliken charges routine in Molpro. Partial charges were averaged over the same configurations as used in the chemical shielding simulations. 
The partial charges of \ce{DMP-} - \ce{Mg^2+} and \ce{DMP-} - \ce{Ca^2+} CIP were averaged over 100 MD snapshots with a \SI{4.5}{\angstrom} hydration shell, five \ce{Ca^2+} configurations were discarded for convergence issues.

For a mapping of the $^{31}P$ chemical shift on geometric parameters, P-O distances of \ce{DMP-} were scanned in a range \SIrange{1.35}{1.60}{\angstrom} for O1/O2-P
non-bridging oxygen atoms ($\Delta d$=\SI{0.01}{\angstrom}) and \SIrange{1.40}{1.80}{\angstrom} for O3/O5-P
bridging oxygen atoms ($\Delta d$=\SI{0.04}{\angstrom}) and $\sigma_{iso}(^{31}$P) 
were evaluated for every geometry (cc-pCVTZ).
Test simulations showed that the influence of the O-P-O angle on $\sigma_{iso}(^{31}$P) is minor which was thus neglected. 
Chemical shift mapping was performed by assigning the instantaneous \ce{DMP-} configuration to the nearest grid point on the 4-D
nuclear shielding map $\sigma_{iso}^{pred}(^{31}$P)($r_{O1-P}, r_{O2-P}, r_{O3-P}, r_{O5-P}$) (82000 points).

\begin{table}
    \caption{Average change of atomic partial charges $\Delta q$ upon CIP formation in \ce{DMP-}-\ce{Mg^2+} and \ce{DMP-}-\ce{Ca^2+} with a \SI{4.5}{\angstrom}  hydration shell on HF and GIAO-DF-LMP2 level of theory.}
    \label{tab:partial_charges_mg_ca}
    \begin{tabular}{lllll}
    \hline
    Atom & \multicolumn{2}{c}{$\Delta  q $ (HF) / $e^-$}  & \multicolumn{2}{c}{$\Delta  q $ (LMP2) / $e^-$} \\
      \hline
       & \ce{Mg^2+} &\ce{Ca^2+} &  \ce{Mg^2+} &\ce{Ca^2+} \\
        \ce{P}    & -0.051 & 0.013  & -0.014 & -0.177    \\
        O1        & 0.154 & 0.056 & 0.138 & 0.110       \\
        O2        & -0.006 & -0.011 & -0.004 & 0.031        \\
        \ce{I^2+} & -0.045 & -0.035 & -0.126 & -0.128  \\
        \hline
    \end{tabular}

\end{table}